\documentclass[%
aip,
showpacs,superscriptaddress,
amsmath,amssymb,
preprint,%
]{revtex4-1}

\usepackage{graphicx}
\usepackage{dcolumn}
\usepackage{bm}

\usepackage[utf8]{inputenc}
\usepackage{eucal}
\usepackage[T1]{fontenc}
\usepackage{amsmath}
\usepackage{mathptmx}
\usepackage{etoolbox}
\usepackage{listings}
\usepackage{enumitem}
\usepackage{diagbox}
\DeclareMathOperator{\sech}{sech}
\usepackage{color,soul}
\usepackage{xcolor}

\definecolor{codegreen}{rgb}{0,0.6,0}
\definecolor{codegray}{rgb}{0.5,0.5,0.5}
\definecolor{codepurple}{rgb}{0.58,0,0.82}
\definecolor{backcolour}{rgb}{0.95,0.92,1.0}

\lstdefinestyle{mystyle}{
	backgroundcolor=\color{backcolour},   
	commentstyle=\color{codegreen},
	keywordstyle=\color{magenta},
	numberstyle=\tiny\color{green},
	stringstyle=\color{codepurple},
	basicstyle=\ttfamily\footnotesize,
	breakatwhitespace=false,         
	breaklines=true,                 
	captionpos=b,                    
	keepspaces=true,                 
	numbers=none,                    
	numbersep=5pt,                  
	showspaces=false,                
	showstringspaces=false,
	showtabs=false,                  
	tabsize=2,
	frame=shadowbox,
	rulecolor=\color{blue}
}

\makeatletter
\def\@email#1#2{%
	\endgroup
	\patchcmd{\titleblock@produce}
	{\frontmatter@RRAPformat}
	{\frontmatter@RRAPformat{\produce@RRAP{*#1\href{mailto:#2}{#2}}}\frontmatter@RRAPformat}
	{}{}
}%
\makeatother
\begin{document}

	\title{Data driven soliton solution of the nonlinear Schr\"odinger equation with certain $\mathcal{PT}$-symmetric potentials via deep learning}

	\author{J. Meiyazhagan}
	\affiliation{Department of Nonlinear Dynamics, Bharathidasan University, Tiruchirappalli 620 024, Tamil Nadu, India.}
	\author{K. Manikandan}%
	\affiliation{Centre for Nonlinear Systems, Chennai Institute of Technology, Chennai, 600 069, Tamil Nadu, India.}
	\author{J. B. Sudharsan}%
	\affiliation{Centre for Nonlinear Systems, Chennai Institute of Technology, Chennai, 600 069, Tamil Nadu, India.}
	\author{M. Senthilvelan}%
	
	\affiliation{Department of Nonlinear Dynamics, Bharathidasan University, Tiruchirappalli 620 024, Tamil Nadu, India.}%
	\email{velan@cnld.bdu.ac.in}
	
	\date{\today}
	
	\begin{abstract}
		We investigate the physics informed neural network method, a deep learning approach, to approximate soliton solution of the nonlinear Schr\"odinger equation with parity time symmetric potentials. We consider three different parity time symmetric potentials, namely Gaussian, periodic and Rosen-Morse potentials. We use physics informed neural network to solve the considered nonlinear partial differential equation with the above three potentials. We compare the predicted result with actual result and analyze the ability of deep learning in solving the considered partial differential equation. We check the ability of deep learning in approximating the soliton solution by taking squared error between real and predicted values. {Further, we examine the factors that affect the performance of the considered deep learning method with different activation functions, namely ReLU, sigmoid and tanh. We also use a new activation function, namely sech which is not used in the field of deep learning and analyze whether this new activation function is suitable for the prediction of soliton solution of nonlinear Schr\"odinger equation for the aforementioned parity time symmetric potentials. In addition to the above, we present how the network's structure and the size of the training data influence the performance of the physics informed neural network. Our results show that the constructed deep learning model successfully approximates the soliton solution of the considered equation with high accuracy.}
	\end{abstract}
	
	\maketitle
	
	\begin{quotation}
	Soliton in complex parity time ($\mathcal{PT}$) symmetric media has attained an undeniable advantage of gain and loss distribution in its dynamics.  Such dynamical studies on solitons have been widely analyzed in mode-locked lasers, ultrashort pulse optics, optical solitons in communication systems and atomic lasers. Solving partial differential equations (PDEs) using deep learning approaches have become popular because the method depends on the optimization techniques. Physics Informed Neural Networks (PINN) are one kind of optimization method which is used to solve a wide class of PDEs including the nonlinear Schr\"odinger (NLS) equation. We utilize the PINN method to approximate soliton solution of the NLS equation with three different $\mathcal{PT}$-symmetric potentials, namely Gaussian, periodic and Rosen-Morse potentials. The PINN method accurately approximates the soliton solution of the considered NLS equation for all three potentials.
	\end{quotation}
	
	\section{Introduction}
	\par For the past four decades, soliton and its applications have been studied in depth in several branches of optics.  In particular, the demand of harnessing the fruitfulness of solitons in nonlinear fiber optics and communication systems have attracted plethora of interests \cite{malomed}. Mathematically, the dynamics of such optical soliton pulses can be described by the nonlinear Schr\"odinger (NLS) equation. By properly managing the dispersion and nonlinearity parameters in the NLS equation one can generate a stable soliton.
	\par It has also been shown that by introducing a proper complex parity-time ($\mathcal{PT}$) symmetric potential in the NLS equation, one can gain more access on the optical soliton pulse propagation. Even though the complex $\mathcal{PT}$-symmetric potential is non-Hermitian in nature, the underlying system admits real eigenspectra \cite{bender1998} and it also supports continuous range of stable optical solitons \cite{musslimani2008optical,kominis2019continuous,zyan}. The dynamical behaviours of $\mathcal{PT}$-symmetric optical solitons have been investigated in many optical experiments and theoretical models \cite{guo2009observation,Ruter,regensburger,Makris,Chen2,wen,hari,
		manikandan2018deformation,manikandan2021nonlinear}. 
	\par Nowadays Machine Learning (ML) and Deep Learning (DL) approaches have become important tools in the prediction task in various fields of physics \cite{Carleo2019,sudhe1,sudhe3,santo1}. In the field of nonlinear dynamics, ML methods have been used for the replication of chaotic attractors \cite{Pathak2017}, prediction of chaotic laser pulses amplitude \cite{Amil2019}, detection of unstable periodic orbits \cite{zhu2019},  chaotic signals separation \cite{Krishnagopal2020}, network classification from symbolic time series \cite{Panday2021}, identification of chimera states \cite{BARMPARIS2020,ganaie2020,kushwaha2020} and also in the study of extreme events \cite{lellep2020, meiysudha1, dibak1, ray2021optimized, meiysudha2}.
		
	\par The rapid growth in the field of DL enables us to solve linear and nonlinear partial differential equations (PDEs) by an approximation technique, namely Physics Informed Neural Network (PINN) which was introduced by Raissi {et al.} \cite{raissi2019physics}. For the past couple of years PINNs have been widely used to solve NLS equation and its generalizations \cite{pu2021data, wang2021data, zhou2021deep, wang2021data2, mo2022data}. In this direction, quite recently, the logarithmic NLS equation with $\mathcal{PT}$-symmetric harmonic potential and Scarf-II potential have been solved through PINN approach \cite{zhou2021solving,li2021solving}. In the present work, we consider NLS equation with three different $\mathcal{PT}$-symmetric potentials, namely Gaussian, periodic and Rosen-Morse potentials and approximate the soliton solution of all three cases with the PINN approach. In our study, we introduce a new activation function, namely $\sech$ and test the ability of this new function by comparing it with the other activation functions that are being used in the literature. To the best of our knowledge, this is the first time wherein the PINN approach is being used to solve the NLS equation with the above said $\mathcal{PT}$-symmetric potentials and this is also the first time $\sech$ is used as an activation function in this approach.
	
	\par We organize our presentation as follows. In Sec. II, we present the methodologies involved in the PINN approach and the general way of solving the considered NLS equation with PINN method. The data driven soliton solution of the NLS equation with all three considered potentials, a comparison with exact solution and the error occurring in this approximation are given in Sec. III. A comparative study on {factors that affect the performance of the PINN is discussed in Sec. IV}. We present our conclusions in Sec. V.
		
	\section{PINN and the NLS equation with $\mathcal{PT}$-symmetric potential}
	\subsection{The scheme of PINN}
	\par Usually, the PINNs have been used for solving nonlinear PDEs that have the general form \cite{raissi2019}
	\begin{equation}
		u_t - \mathcal{N}[u(x,t);\lambda] = 0, \quad x \in \Omega,\quad t\in[0,T]. \label{main_eq}
	\end{equation}
	In this work, we consider the complex nonlinear PDEs with the following initial and boundary conditions, that is
	\begin{equation}
		\begin{cases}\label{pdeibc}
			iu_t=\mathcal{N}[u(x,t);\lambda_0], \quad &x \in \Omega,\quad t\in[0,T],\\
			I[u(x,t)]|_{t=0}=u_I(x), \quad &x\in\Omega\; (\textrm{initial condition}),\\
			B[u(x,t)]|_{x\in\partial \Omega}=u_B(t), \quad &t\in [0,t]\;(\textrm{boundary conditions}),
		\end{cases}
	\end{equation}
	where $u(x,t)$ is the solution of the PDE,  $\mathcal{N}[.,\lambda]$ is the combination of linear and nonlinear operators which are parametrized by the initial vector $\lambda_0$, $[0,T]$ represents the lower and upper boundary of the time variable $t$, $\Omega$ and $\partial\Omega$ denote the spatial variable range and the boundary of that domain respectively, $I$ and $B$ are operators corresponding to initial and boundary values, $I[u(x,t)]|_{t=0}=u_I(x)$ and $B[u(x,t)]|_{x\in\partial \Omega}=u_B(t)$ respectively represent the initial and boundary conditions. We define a complex-valued physics model $f(x,t)$ as follows
	\begin{equation}
		f(x,t) := iu_t-\mathcal{N}[u;\lambda_0].
	\end{equation}
	\par We can differentiate the latent solution $u(x,t)$ with respect to time variable $t$ and spatial variable $x$ using the derivative technique, namely Automatic Differentiation (AD) \cite{baydin2018,margossian2019review} based on the chain rule which is used to make Back Propagation (BP) \cite{rumelhart1986} in Artificial Neural Networks (ANN). For the implementation of BP, AD and other optimization steps involved in the complex-valued PINN, we use Tensorflow \cite{abadi2016}, which is a well known open-source software library used for AD and DL computations. We use four different kinds of activation functions for the activation of neurons in the ANN (a comparative study on the activation functions is given in Sec.~\ref{a_fn_sec}). However, for the main study, we choose $\tanh$ as the nonlinear activation function which is being used in the current literature in the form \cite{raissi2019}
	\begin{equation}
		Z_l = \tanh(w_l.Z_{l-1}+b_l), \qquad l = 1,2,3,...n, \label{tanh}
	\end{equation}
	where $w_l$ is the dim($Z_l$)$\times$dim($Z_{l-1}$) weight matrix and $b_l$ is the dim($Z_l$) bias vector. We define the loss function for the training process as
	\begin{equation}
		\begin{aligned}[b]\label{loss}
			L_{Train} = &\frac{1}{N_I}\sum_{j=1}^{N_I} \left|I[u(x_I^j,t)]|_{t=0}-u_I(x_I^j)\right|^2\\&+\frac{1}{N_B}\sum_{j=1}^{N_B} \left|B[u(x_B,t_B^j)]|_{x_B\in\partial D}-u_B(t_B^j)\right|^2\\
			&+\frac{1}{N_C}\sum_{j=1}^{N_C}\left|f(x_C^j,t^j_C)\right|^2,
		\end{aligned}
	\end{equation}
	where $\left\{x_I^j,u_I^j\right\}_{j=1}^{N_I}$ and $\left\{t_B^j,u_B^j\right\}_{j=1}^{N_B}$ are respectively represent the initial and boundary conditions and $\left\{x_C^j, t_C^j, f(x_C^j,t_C^j)\right\}_{j=1}^{N_C}$ denote the collocation points of $f(x,t)$. We create the sample points using Latin Hypercube Sampling (LHS) algorithm \cite{stein1987} and the optimization for the loss function by Limited memory Broyden–Fletcher–Goldfarb–Shanno (L-BFGS) algorithm \cite{liu1989}.
	\par The major steps involved in solving the PDE \eqref{main_eq}, with initial and boundary conditions \eqref{pdeibc}, using the PINN method, are given below:
	
	\begin{enumerate}[label=(\roman*)]
		\item Defining the {structure of ANN which is described by fixed number of layers and fixed number of neurons}.
		\item Preparing three training sets, namely (i) the initial condition set, (ii) boundary conditions sets and (iii) the random collocation points using the LHS technique \cite{stein1987}.
		\item Getting the training loss function $L_{Train}$ given in \eqref{loss} by adding weighted $\mathbb{L}^2$-norm errors of the initial, boundary condition residuals and $f(x,t)$.
		\item Train the ANN in order to {get suitable values of  $\{\mathbf{\hat{w}},\mathbf{\hat{b}}\}$ to minimize the $L_{Train}$} using the L-BFGS algorithm.
	\end{enumerate}
	Using these four steps we approximate the solution of the considered PDE.
	\subsection{PINN method for NLS equation with $\bm{\mathcal{PT}}$-symmetric potential}
	\par We consider NLS equation with a $\mathcal{PT}$-symmetric potential in the form,
	\begin{equation}
		i\psi_t+\psi_{xx}+P(x)\psi+\sigma|\psi|^2\psi=0, \label{nls_main}
	\end{equation}
	where $\psi=\psi(x,t)$ is a complex field, $\sigma$ is the nonlinear coefficient corresponding to focusing and defocusing interactions and $P(x)$ is the $\mathcal{PT}$-symmetric potential which has the form,
	\begin{equation}
		P(x) = [V(x)+iW(x)],
	\end{equation}
	where $V(x)$ and $W(x)$ are real and imaginary parts of the $\mathcal{PT}$-symmetric potential and they should satisfy the following two conditions:
		\begin{equation}
			V(-x) = V(x), \qquad	W(-x) = -W(x). \label{con1}
		\end{equation}
	\par In this work, to obtain the soliton solution of \eqref{nls_main} using the above mentioned PINN method \cite{raissi2019}, we define the equation, initial and boundary conditions respectively as follows:
	\begin{equation}
		i\psi_t = -\psi_{xx}-P(x)\psi-\sigma|\psi|^2\psi, \quad x\in(-L,L), \quad t\in(0,T),\label{nls2}
	\end{equation}
	\begin{subequations}
		\begin{equation}
			\psi(x,0)=\psi_0(x), \quad x\in [-L,L],
		\end{equation}
		\begin{equation}
			\psi(-L,t) = \psi(L,t), \quad t \in [0,T].
		\end{equation}
	\end{subequations}
	\par
\begin{figure*}[!ht]
		\includegraphics[width=0.9\linewidth]{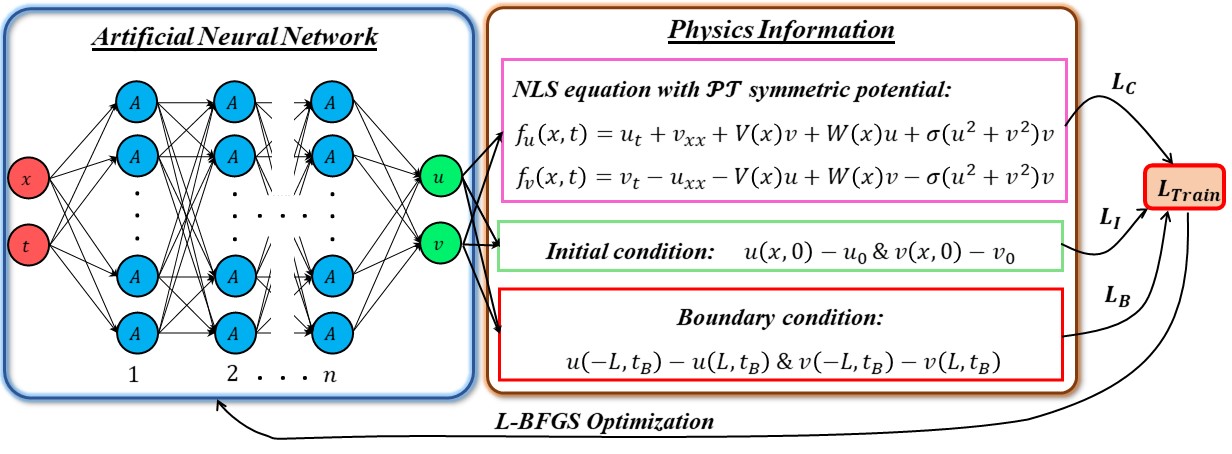}
		\caption{\label{fig:pinn} Schematic diagram of PINN. The left panel shows the input and output layers with two neurons and $n$ hidden layers. Each neuron is activated by the activation function $A$. The right panel corresponds to the physics information of the PINN given as a loss function of the optimization problem.}
	\end{figure*}

 Since the solution $\psi(x,t)$ of Eq. \eqref{nls2} is complex, we consider it in the form $\psi(x,t)=u(x,t)+iv(x,t)$, where $u(x,t)$ and $v(x,t)$ are two real functions denoting real and imaginary parts of the solution $\psi$, respectively. Now we use the associated complex-valued PINN as $f(x,t)=if_u(x,t)-f_v(x,t)$ with $-f_v(x,t)$ and $f_u(x,t)$ being the real and imaginary parts of $f(x,t)$ respectively. The explicit form of the functions read
	\begin{subequations}
		\begin{equation}
			f(x,t) = i\psi_t+\psi_{xx}+[V(x)+iW(x)]\psi+\sigma|\psi|^2\psi,
		\end{equation}
		\begin{equation}
			f_u(x,t) = u_t+v_{xx}+V(x)v+W(x)u+\sigma (u^2+v^2)v,
		\end{equation}
		\begin{equation}
			f_v(x,t) = v_t-u_{xx}-V(x)u+W(x)v-\sigma (u^2+v^2)u,
		\end{equation}
	\end{subequations}
	and using this, we approximate $\psi(x,t)$ by a complex-valued deep neural network. The shared parameters between $\psi(x,t)$ and $f(x,t)$ can be trained by minimizing $L_{Train}$ which is the combination of three mean squared errors as given below
	\begin{equation}
		L_{Train} = L_I + L_B + L_C,
	\end{equation}
	where the mean squared errors are taken in the form
	\begin{equation}
		\begin{aligned}[b]\label{losses}
			L_I = &\frac{1}{N_I}\sum_{j=1}^{N_I} \left(\left|u(x_I^j,0)-u_0^j\right|^2+\left|v(x_I^j,0)-v_0^j\right|^2\right),\\
			L_B = &\frac{1}{N_B}\sum_{j=1}^{N_B} \left(\left|u(-L,t_B^j)-u(L,t_B^j)\right|^2+\left|v(-L,t_B^j)-v(L,t_B^j)\right|^2\right),\\
			L_C = &\frac{1}{N_C}\sum_{j=1}^{N_C}\left(\left|f_u(x_C^j,t^j_C)\right|^2+\left|f_v(x_C^j,t^j_C)\right|^2\right),
		\end{aligned}
	\end{equation}
	with $\{x_I^j,u_0^j,v_0^j\}_{j=1}^{N_I}$ denote the initial data, $\{t_B^j,u(\pm L,t_B^j), v(\pm L,t_B^j)\}_{j=1}^{N_B}$ denote the boundary data and $\{x_C^j,t_C^j,f_u(x_C^j,t_C^j),f_v(x_C^j,t_C^j)\}_{j=1}^{N_C}$ denote the collocation points on $f(x,t)$. The losses $L_I$, $L_B$ and $L_C$ respectively represent the $\mathbb{L}^2$-norm error in initial, boundary and inside the spatio-temporal regime. Figure \ref{fig:pinn} shows the schematic diagram of the PINN. The left panel of the figure corresponds to the ANN where we have two input neurons for space and time and two output neurons for real and imaginary parts of the solution. The right panel shows the physics information which we give as a form of training loss function $L_{Train}$. 
	
	\section{Data driven solutions of the NLS equation with $\mathcal{PT}$-symmetric potentials}
	\par In this section, we present the outcomes of DL while solving the focusing ($\sigma=1$) NLS equation with soliton solution with three $\mathcal{PT}$-symmetric potentials, namely (i) Gaussian, (ii) periodic and (iii) Rosen-Morse potential. To make the PINN to solve the considered problem, we are in need of training set data. The training set consists of $N_I = 50$ data points on initial conditions, $N_B=100$ data points on the periodic boundary conditions {(50 on the upper boundary and another 50 on the lower boundary)} and $N_C=20000$ collocation points which are chosen randomly using the LHS \cite{stein1987} method. We choose a 6-layer ANN in which the first and the last layer having two neurons which are used for input $(x,t)$ and output $(u(x,t),v(x,t))$. The other 4 hidden layers have 100 units of neurons each. The hyperbolic tangent function given in Eq. \eqref{tanh} is used for the activation of the neurons. The space and time interval are taken as $L=10$ and  $T=5$ respectively. So the limit for spatial and time points are $[-10,10]$  and $[0,5]$. The PINN model has been run for 40,000 optimization steps to minimize the loss function $L_{Train}$. {To verify the outcome of PINN we compare the solutions obtained from PINN with the numerical solutions. To generate the latter data we use Fourier spectral method {\cite{yang2010}} with a special Fourier discretization with 256 space modes and a fourth-order explicit Runge-Kutta temporal integrator with 201 points at the same space/time interval to solve the NLS equation {\eqref{nls2}}. So $\psi(x,t)$ is a $256\times 201$ matrix. We note here that the solution obtained using the above said numerical method is just to access the accuracy of the PINN solution. Training of the PINN itself does not require a numerical solution.}
	
	The above setup has been considered same for all three $\mathcal{PT}$-symmetric potentials throughout this work except the activation function which we change in each case since it will be used to study the influence in the accuracy of solving the NLS equation with $\mathcal{PT}$-symmetric potential using PINN method.
	\subsection{NLS equation with $\mathcal{PT}$-symmetric Gaussian potential}
	\par
\begin{figure}[!ht]
		\includegraphics[width=0.5\linewidth]{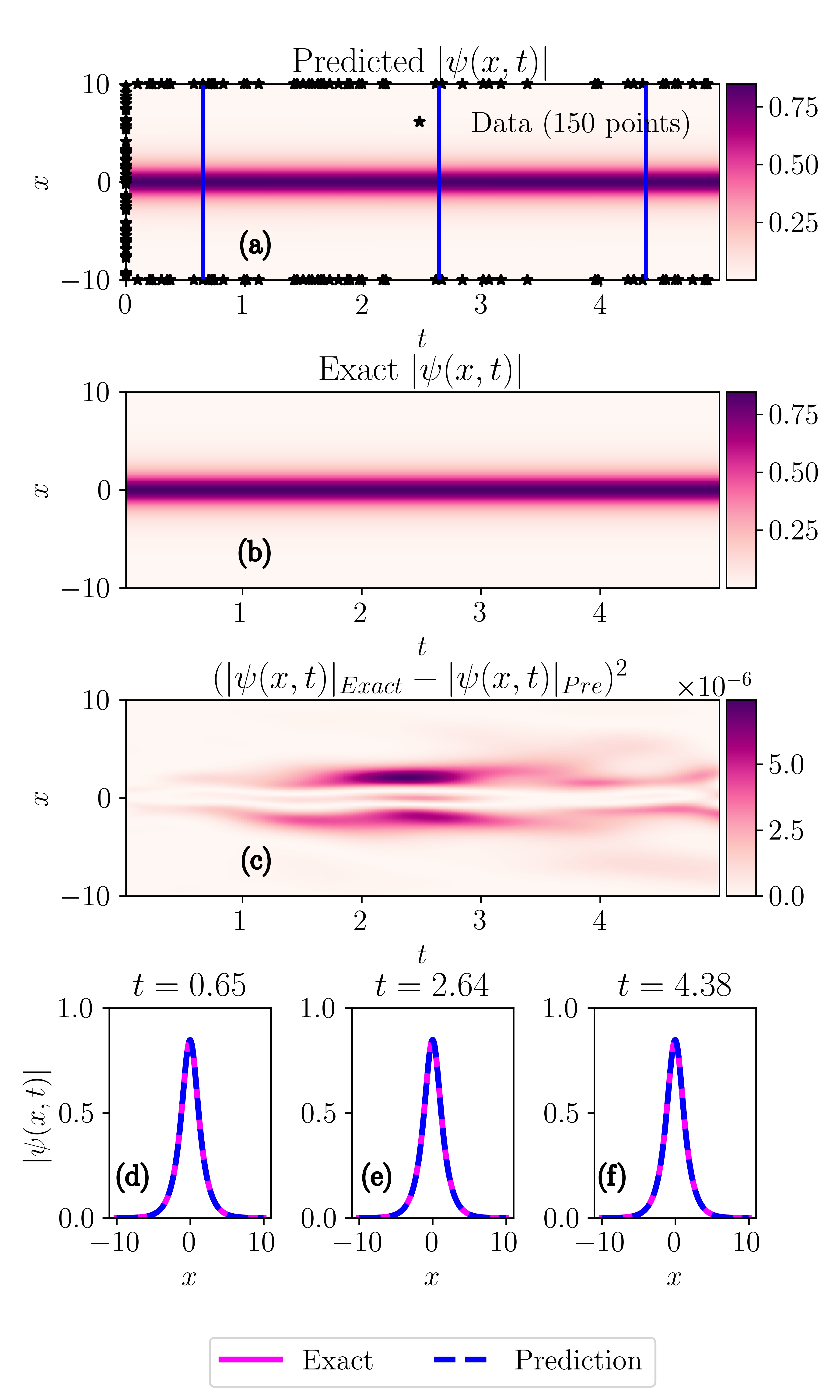}
		\caption{\label{fig:gauss_main} Results of PINN in approximating soliton solution of NLS equation \eqref{nls2} with Gaussian potential \eqref{gauss}. (a) representing the predicted values of $|\psi(x,t)|$ by PINN. The stars in Fig. (a) representing randomly selected data points on initial and boundary conditions. (b) showing the exact values of $|\psi(x,t)|$. (c) corresponding to the squared error values between predicted and exact results. (d)-(f) a comparison of approximation done by PINN in finding soliton solution at particular time instants $t=0.65$, $t=2.64$ and $t=4.38$}
	\end{figure}

 To begin, we consider the NLS equation \eqref{nls2} with $\mathcal{PT}$-symmetric Gaussian potential \cite{hu2011}
	\begin{equation}
		P(x) = V(x) + i W(x) = e^{-x^2}+iW_0 xe^{-x^2}, \label{gauss}
	\end{equation}
	where $W_0$ is the strength of the imaginary part with the value 0.1. The real $(V(x))$ and the imaginary $(W(x))$ parts of the potential $P(x)$ given in \eqref{gauss} satisfy the conditions given in \eqref{con1}. {The Gaussian profile is taken as the initial profile to solve the NLS equation {\eqref{nls2}} with the above potential {\eqref{gauss}}.} After training the PINN with the above mentioned setup with the Gaussian potential, the approximated solution is shown in Fig. \ref{fig:gauss_main}. 
	Figures \ref{fig:gauss_main} (a) and (b) respectively represent the predicted and exact magnitude of the soliton solution $|\psi(x,t)|=\sqrt{u^2(x,t)+v^2(x,t)} $ for the NLS equation \eqref{nls2} with $\mathcal{PT}$-symmetric Gaussian potential \eqref{gauss}. The star markers in Fig. \ref{fig:gauss_main} (a) denote the randomly chosen data points on the initial (50 points) and boundary (100 points) conditions. Figure \ref{fig:gauss_main} (c) shows the value of squared errors between predicted and the exact values of the solution. From Fig. \ref{fig:gauss_main} (a) we can see that the predicted soliton solution of the NLS equation \eqref{nls2} with potential \eqref{gauss} is similar to that of the exact solution shown in Fig. \ref{fig:gauss_main} (b). To examine the error between these two solutions, we plot the squared error of them in Fig. \ref{fig:gauss_main} (c). This figure infers that the error value between predicted and the exact solutions are in the order of $10^{-6}$. The relative $\mathbb{L}^2$-norm errors of $u(x,t)$, $v(x,t)$ and $\psi(x,t)$ respectively are $2.1856 \times 10^{-2}$, $2.8822\times 10^{-2}$ and $3.1912\times 10^{-3}$. These results infer that the considered PINN is enabled to approximate the soliton solution of the NLS equation with considered Gaussian potential with low error values. Figures \ref{fig:gauss_main} (d)-(f) show the comparisons of exact and the predicted soliton solution at different times, say $t=0.65$, $t=2.64$ and $t=4.38$. The predicted solitons at different time instants are fitted well with the exact soliton solutions. This also confirms the ability of PINN in solving the NLS equation for the given Gaussian $\mathcal{PT}$-symmetric potential.

	\subsection{NLS equation with $\mathcal{PT}$-symmetric periodic potential}
	\par
\begin{figure}[!ht]
		\includegraphics[width=0.5\linewidth]{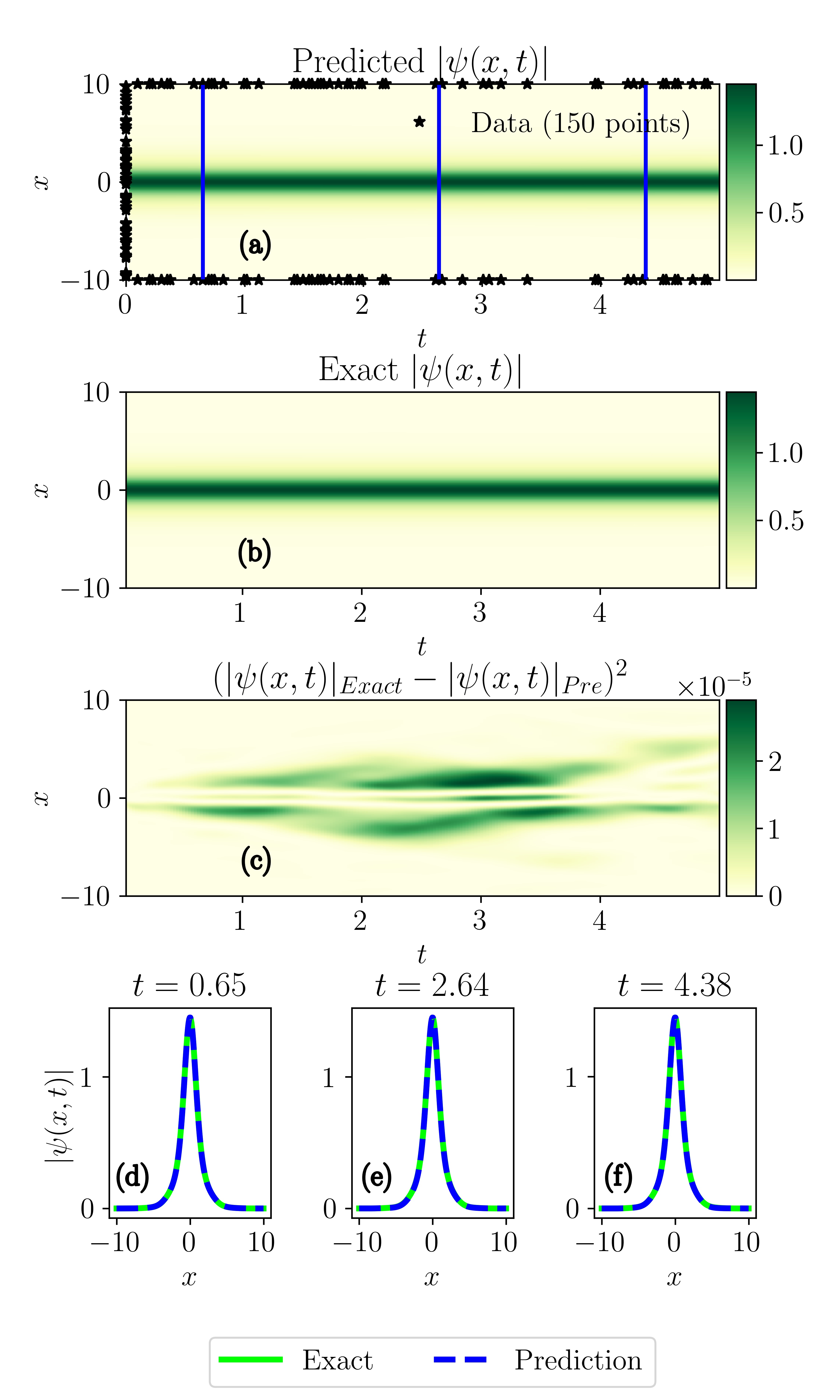}
		\caption{\label{fig:periodic_main} Results of PINN in approximating soliton solution of NLS equation \eqref{nls2} with periodic potential \eqref{periodic}. (a) representing the predicted values of $|\psi(x,t)|$ by PINN. The stars in Fig. (a) representing randomly selected data points on initial and boundary conditions. (b) showing the exact values of $|\psi(x,t)|$. (c) corresponding to the squared error values between predicted and exact results. (d)-(f) a comparison of approximation done by PINN in finding soliton solution at particular time instants $t=0.65$, $t=2.64$ and $t=4.38$}
	\end{figure}

 Let us now consider the potential $P(x)$ in \eqref{nls2} in the form \cite{musslimani2008optical}
	\begin{equation}
		P(x)=V(x)+iW(x) = \cos^2 x+i W_0\sin 2x, \label{periodic}
	\end{equation}
	where the value of strength of the imaginary part is $W_0=0.45$. Since the potential $P(x)$ is $\mathcal{PT}$-symmetric, the real ($V(x)$) and imaginary ($W(x)$) parts satisfy the conditions mentioned in \eqref{con1}. {Here also the Gaussian profile is considered as the initial profile to solve the NLS equation {\eqref{nls2}} with $\mathcal{PT}$-symmetric periodic potential {\eqref{periodic}}.} After the initial common setup made for training the PINN, as mentioned before, we obtain the soliton solution from PINN. The obtained results are reported in Fig.~\ref{fig:periodic_main}.
	The predicted and the exact magnitude of the soliton solution $|\psi(x,t)|$ obtained are shown in Figs.~\ref{fig:periodic_main} (a) and \ref{fig:periodic_main} (b) respectively. The data points which are randomly chosen on the initial and boundary conditions for the purpose of training are denoted as stars in Fig.~\ref{fig:periodic_main} (a). The squared error values between predicted and the exact solutions are shown in Fig.~\ref{fig:periodic_main} (c). From Fig.~\ref{fig:periodic_main} (c), we can see that the error values are in the order of $10^{-5}$ which confirm that our constructed PINN model succeeds in approximating the soliton solution of the NLS equation with $\mathcal{PT}$-symmetric periodic potential \eqref{periodic} with high accuracy. Further, to check the correctness of the solution we plot the solution $|\psi(x,t)|$ at different instants of time, say $t=0.65$, $t=2.64$ and $t=4.38$ in Figs.~\ref{fig:periodic_main} (d), (e) and (f) respectively and these figures also confirm that the solution obtained through PINN is accurate since the exact and the predicted solutions coincide with each other. The relative $\mathbb{L}^2$-norm error values in $u(x,t)$, $v(x,t)$ and $\psi(x,t)$ in this case are found to be $5.0925\times 10^{-2}$, $4.9897\times 10^{-2}$ and $4.272\times 10^{-3}$.

	\subsection{NLS equation with $\mathcal{PT}$-symmetric Rosen-Morse potential}
	\par
\begin{figure}[!ht]
		\includegraphics[width=0.5\linewidth]{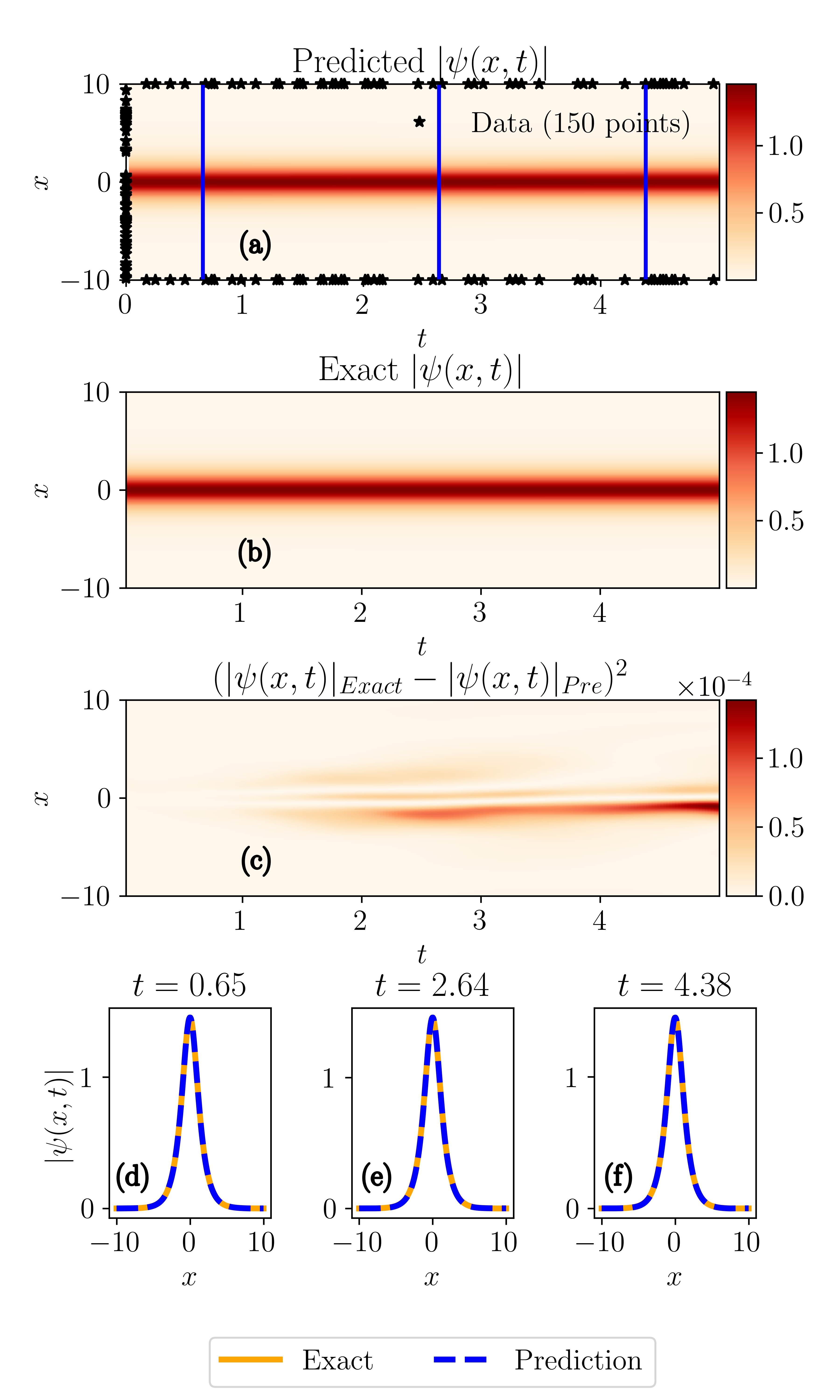}
		\caption{\label{fig:rosen_main} Results of PINN in approximating soliton solution of NLS equation \eqref{nls2} with Rosen-Morse potential \eqref{rose}. (a) representing the predicted values of $|\psi(x,t)|$ by PINN. The stars in Fig. (a) representing randomly selected data points on initial and boundary conditions. (b) showing the exact values of $|\psi(x,t)|$. (c) corresponding to the squared error values between predicted and exact results. (d)-(f) a comparison of approximation done by PINN in finding soliton solution at particular time instants $t=0.65$, $t=2.64$ and $t=4.38$}
	\end{figure}

 Next, we consider another $\mathcal{PT}$-symmetric potential, namely Rosen-Morse potential which is given by\cite{midya2013}
	
	\begin{equation}
		P(x)=V(x)+iW(x) =  -a(a+1)\sech^2 x+i\; 2b\tanh x, \label{rose}
	\end{equation}
	where $a$ and $b$ are parameters which we will take as $0.1$ and $0.03$. The potential considered in \eqref{rose} also satisfies the conditions given in \eqref{con1}. {We take the initial profile in the form {\cite{midya2013}}}
	\begin{equation}
		\psi(x) = \sqrt{a^2+a+2} \sech x e^{ibx}, \label{ini_rosean}
	\end{equation}
	{which satisfies the stationary part of {\eqref{nls2}}}. We consider the same preliminary setup and train the PINN as in the case of Gaussian and periodic potentials and obtain the outcome which we present in Fig.~\ref{fig:rosen_main}. Figures~\ref{fig:rosen_main} (a) and (b) respectively correspond to the predicted and exact magnitudes of the soliton solution of the NLS equation \eqref{nls2} with Rosen-Morse potential \eqref{rose}. The stars at the boundary of Fig.~\ref{fig:rosen_main} (a) denote the data points taken in the initial and boundary conditions. The squared error values between exact and the predicted ones are in the order of $10^{-4}$ which is reported in Fig.~\ref{fig:rosen_main} (c). The Figs.~\ref{fig:rosen_main} (a)-(c) reveal that PINN succeeds in approximating the soliton solution of the Rosen-Morse potential as well. The relative $\mathbb{L}^2$-norm error values of $u(x,t)$, $v(x,t)$ and $\psi(x,t)$ for this case are found to be $3.7277\times 10^{-2}$, $3.2468\times 10^{-2}$ and $5.9112\times 10^{-3}$. In Figs.~\ref{fig:rosen_main} (d)-(f), the exact and the predicted solitons are plotted one over the other at different time instants, say for example $t=0.65$, $t=2.64$ and $t=4.38$ in order to check whether the predicted result is accurate or not. From these figures we can see that the exact and the predicted solutions fit well one over the other indicating that the predicted result is accurate.
	\subsection{Non-stationary solution of NLS equation with $\mathcal{PT}$-symmetric Rosen-Morse potential}
	\par
\begin{figure}[!ht]
		\includegraphics[width=0.5\linewidth]{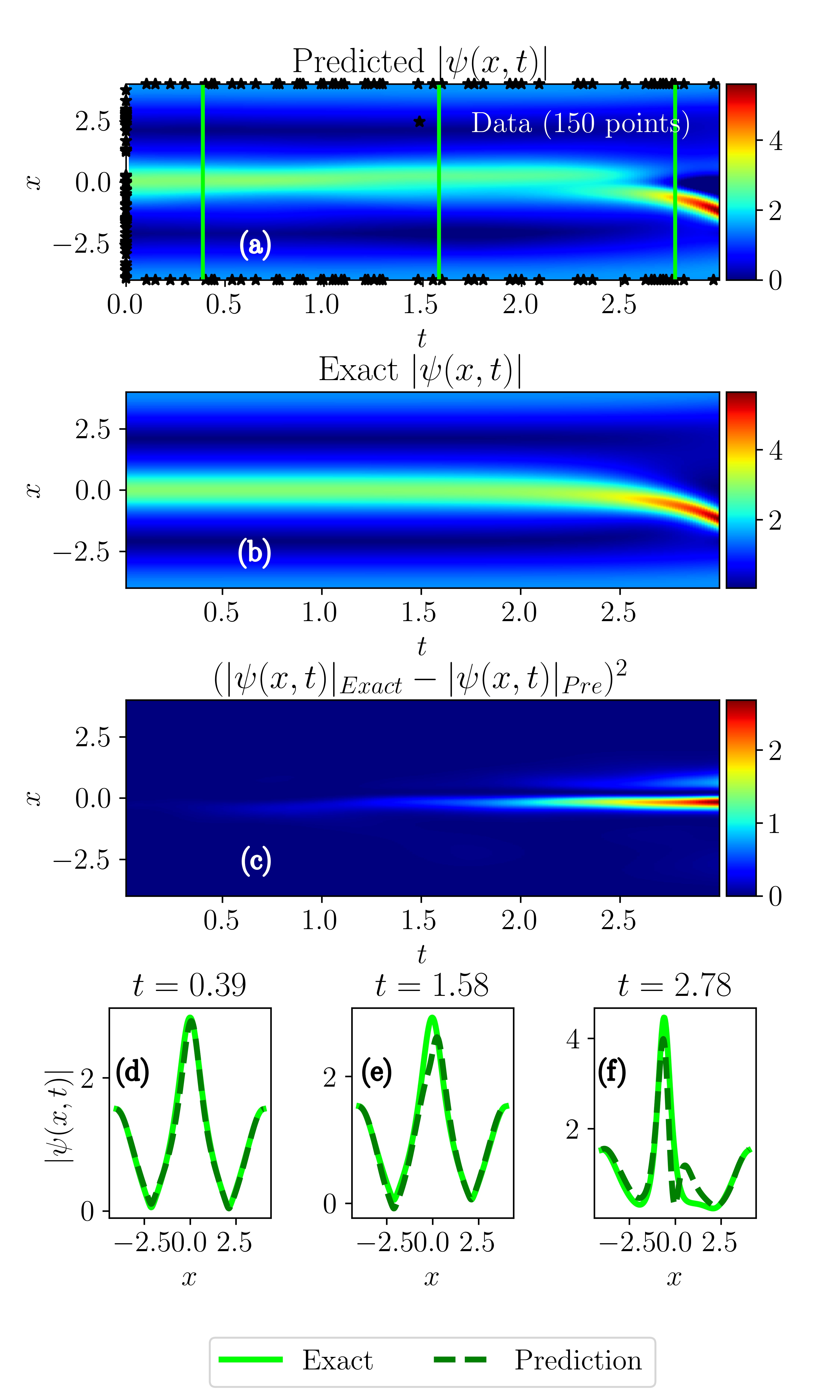}
		\caption{\label{fig:rosen_non_stat} Results of PINN in approximating non-stationary solution of NLS equation \eqref{nls2} with Rosen-Morse potential \eqref{periodic}. (a) representing the predicted values of $|\psi(x,t)|$ by PINN. The stars in Fig. (a) representing randomly selected data points on initial and boundary conditions. (b) showing the exact values of $|\psi(x,t)|$. (c) corresponding to the squared error values between predicted and exact results. (d)-(f) a comparison of approximation done by PINN in finding soliton solution at particular time instants $t=0.39$, $t=1.58$ and $t=2.78$}
	\end{figure}

 {Finally, we analyze the ability of PINN in predicting non-stationary solutions of the NLS equation. Let us consider an initial profile of a non-stationary solution to the NLS equation for the potential {\eqref{rose}} in the form}
	\begin{equation}
		\psi(x) = \sqrt{a^2+1} \sech x e^{-ibx}, \label{ini_rosean_non}
	\end{equation}
	{which does not satisfy the stationary part of {\eqref{nls2}}. Let us fix the parameters as $a=1.75$ and $b=0.35$. The obtained results after training the PINN with the same preliminary setup considered earlier are reported in Fig.~{\ref{fig:rosen_non_stat}}. The predicted and exact magnitudes of $\psi(x,t)$ are presented respectively in Figs.~{\ref{fig:rosen_non_stat}} (a) and {\ref{fig:rosen_non_stat}} (b). From these two figures we observe that PINN successfully predict the non-stationary solution as well.}
 	{We have examined the error values between predicted and exact results and plot the outcome in Fig.~{\ref{fig:rosen_non_stat}} (c). We come across the relative $\mathbb{L}^2$-norm error values of $u(x,t)$, $v(x,t)$ and $\psi(x,t)$ for this case as $5.4182\times 10^{-1}$, $5.5958\times 10^{-1}$ and $2.8065\times 10^{-1}$ respectively. To verify the obtained solution we also plot the solution at different time instants, say $t=0.39$, $t=1.58$ and $t=2.78$ in Figs.~{\ref{fig:rosen_non_stat}} (d), (e) and (f) respectively. The exact and predicted magnitudes of the predicted solution are fitted well in figures (d) and (e) but in Fig.~\ref{fig:rosen_non_stat}~(f) we can observe that the magnitude of the solution not fitted well with the exact one. This is due to the time-dependent nature of the solution. Our investigations reveal that one can solve the considered problem with less accuracy using PINN.}
	\section{Factors affecting the performance of PINN}
	\subsection{Effect of activation functions} \label{a_fn_sec}
	\par
\begin{figure}[!ht]
		\includegraphics[width=0.5\linewidth]{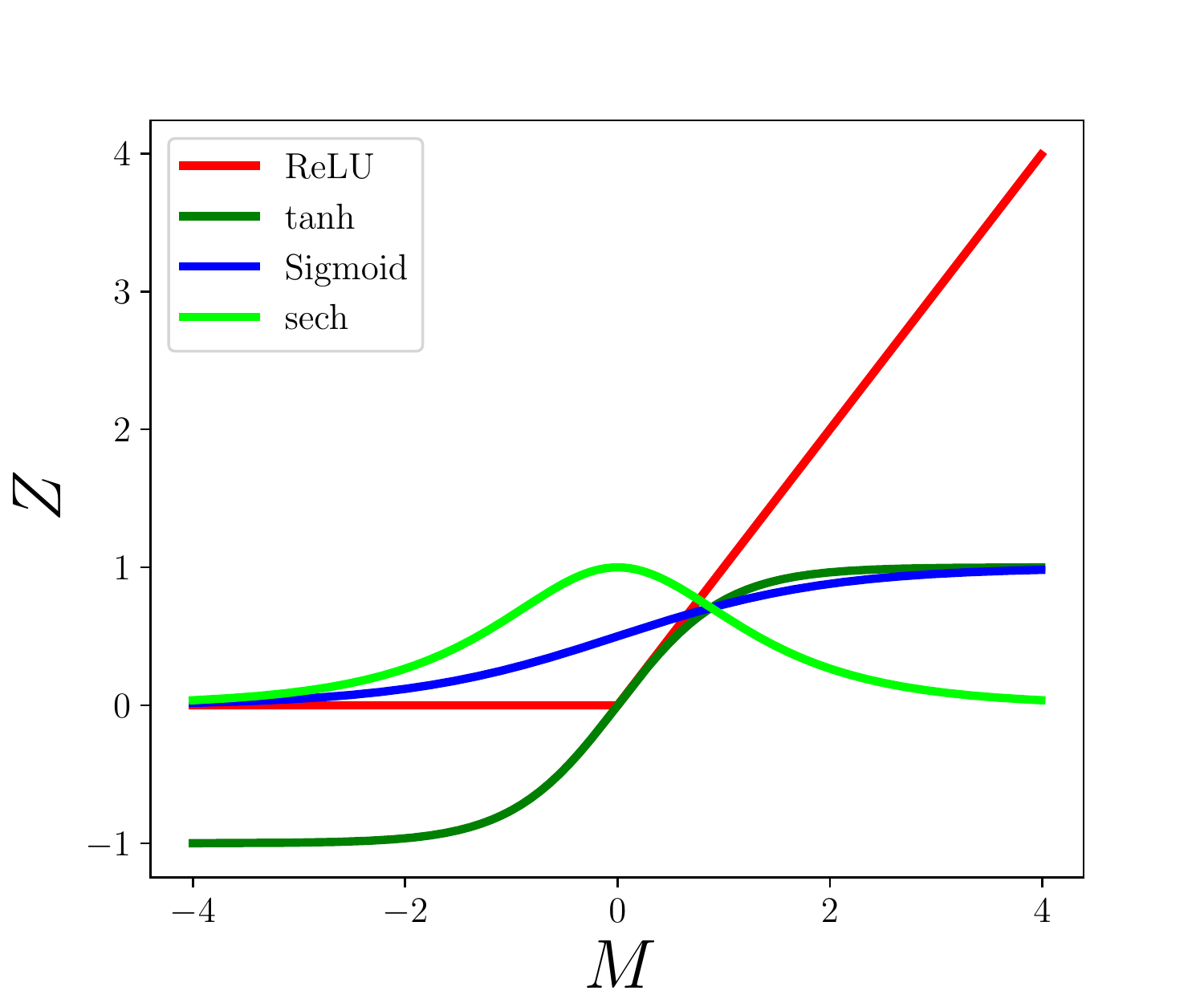}
		\caption{\label{fig:afun} Different kinds of activation functions}
	\end{figure}

 In our main study, we have chosen $\tanh$ as the activation function (see Eq.~\eqref{tanh}) because  it gives us the solution with a low error value. To study the effect of other activation functions in approximating the soliton solution of the $\mathcal{PT}$-symmetric potentials, we consider three other functions, namely Rectified Linear Unit (ReLU), sigmoid and $\sech$. Since the problem under consideration is approximating the soliton solution of the NLS equation with various $\mathcal{PT}$-symmetric potentials, we intend to use a new activation function, namely $\sech$, which has not been used in the field of DL. We consider the general form of the activation functions as given below
	
	\begin{subequations}
		\begin{equation}
		\textrm{(i)}\;	Z_j = ReLU(M) = max(0, M),
		\end{equation}
		\begin{equation}
		\textrm{(ii)}\;Z_j = sigmoid(M) = \frac{1}{1+e^{-M}},
		\end{equation}
		\begin{equation}
		\textrm{(iii)}\;Z_j = \sech(M),\qquad\qquad\qquad
		\end{equation}
		\begin{equation}
		\textrm{(iv)}\;Z_j = \tanh(M),\qquad\qquad\qquad
		\end{equation}
	\end{subequations}
	where $M = w_j.Z_{j-1}+b_j$ as taken in \eqref{tanh}. The functionality of each activation function can be visualized with the help of Fig.~\ref{fig:afun}.
	
	\begin{figure*}[!ht]
		\includegraphics[width=0.85\linewidth]{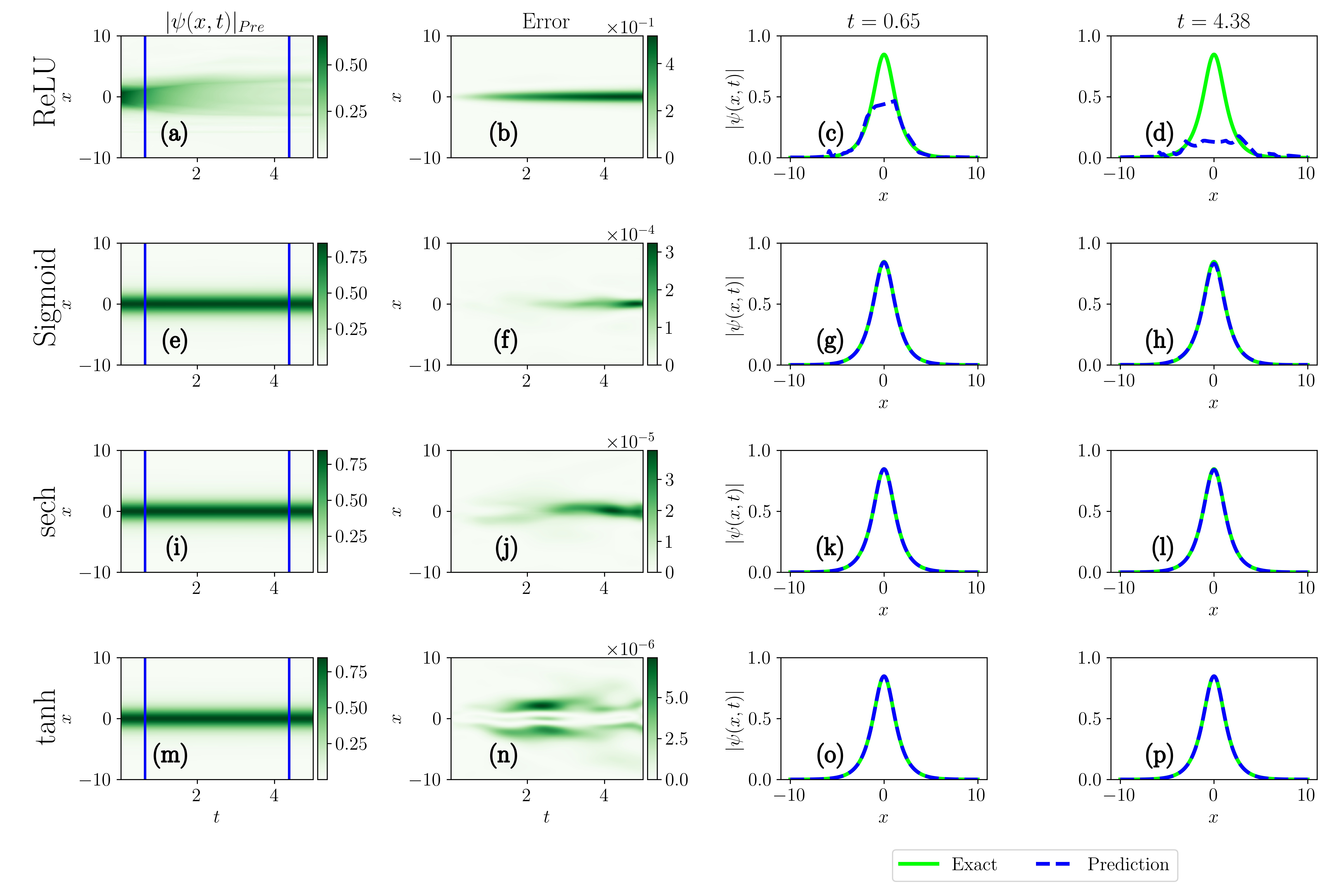}
		\caption{\label{fig:afun_gauss} PINN results for the Gaussian potential case with different activation functions. Rows one to four, respectively, represent the activation functions ReLU, sigmoid, sech and tanh.  {Figs. (a), (e), (i) and (m) represent the predicted magnitude of soliton solutions. (b), (f), (j) and (n) represent the error values in magnitude of soliton solutions. Figs. (c), (g), (k) and (o) in third column and Figs. (d), (h), (l) and (p) in fourth column correspond to the soliton solution at particular time instants $t=0.65$ and $t=4.38$ respectively}}
	\end{figure*}
	
	\par The predicted values and the error values in the case of NLS equation with $\mathcal{PT}$-symmetric Gaussian potential in approximating the soliton solution with all four different activation functions are reported in Fig.~\ref{fig:afun_gauss}. Figure \ref{fig:afun_gauss} (a) reveals that the function ReLU fails in approximating the soliton solution. The squared error value between predicted and the actual magnitude of soliton solution comes out in the order of $10^{-1}$ only (Fig.~\ref{fig:afun_gauss} (b)). The comparison between predicted and the exact solutions at two other instants of time, say at $t=0.65$ and $t=4.38$ are presented in Figs.~\ref{fig:afun_gauss} (c) and (d) which also confirms that by using ReLU as the activation function, a good approximation cannot be obtained for Gaussian potential. The results for the other three activation functions, namely sigmoid, sech and tanh are plotted in Figs.~\ref{fig:afun_gauss} (e)-(h), (i)-(l) and (m)-(p) respectively. From the outcome, we can infer that the prediction done by PINN with $\tanh$ as activation function gives an accurate result when compared with the other three. As far as the Gaussian potential is concerned, we come across squared error values that are in the order of $10^{-1}, 10^{-4}, 10^{-5}$ and $10^{-6}$ for the ReLU, sigmoid, $\sech$ and $\tanh$ activation functions respectively. 
	
	\begin{figure*}[!ht]
		\includegraphics[width=0.8\linewidth]{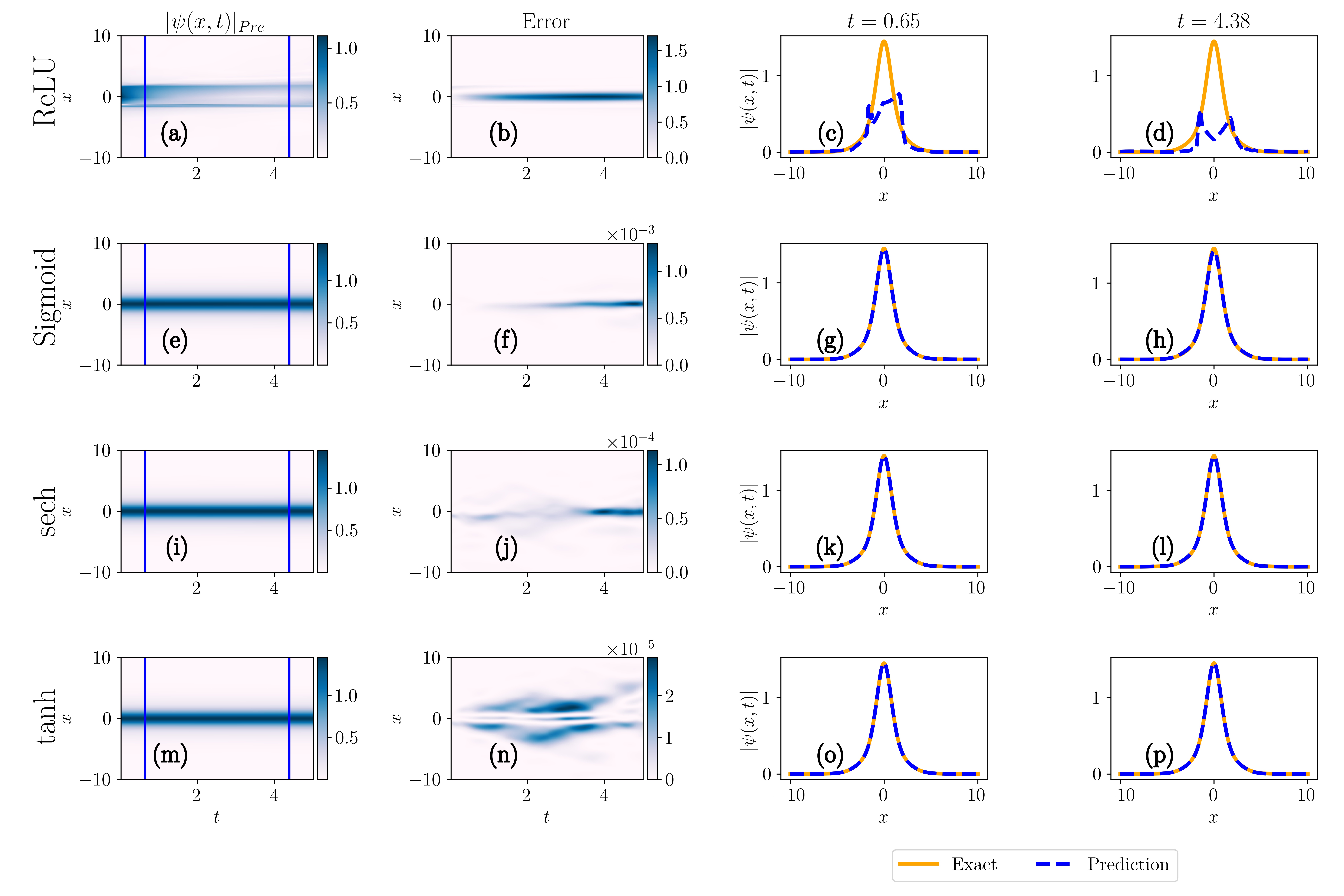}
		\caption{\label{fig:afun_periodic} PINN results for the periodic potential case with different activation functions. Rows one to four, respectively, represent the activation functions ReLU, sigmoid, sech and tanh. {Fig. (a), (e), (i) and (m) represent the predicted magnitude of soliton solutions. Figs. (b), (f), (j) and (n) represent the error values in magnitude of soliton solutions. Figs. (c), (g), (k) and (o) in third column and Figs. (d), (h), (l) and (p) in fourth column correspond to the soliton solution at particular time instants $t=0.65$ and $t=4.38$ respectively.}}
	\end{figure*}
	
	\par The results coming out from PINN with four different activation functions for the $\mathcal{PT}$-symmetric periodic potential is reported in Fig.~\ref{fig:afun_periodic}. The rows one to four represent the results coming out from PINN with the activation functions ReLU, sigmoid, sech and tanh respectively. From Figs.~\ref{fig:afun_periodic} (c) and (d) we can see that the approximation done by PINN with ReLU as an activation function is not fitting well with the original result. But in the case of sech, the prediction is better while comparing with the prediction done by ReLU and sigmoid functions since the squared error value comes out less. From the plots given in the second column of Fig.~\ref{fig:afun_periodic} we infer that the squared error values of the cases ReLU, sigmoid, $\sech$ and $\tanh$ are in the order of $10^{0}, 10^{-3}, 10^{-4}$ and $10^{-5}$ respectively. 
	\begin{figure*}[!ht]
		\includegraphics[width=0.8\linewidth]{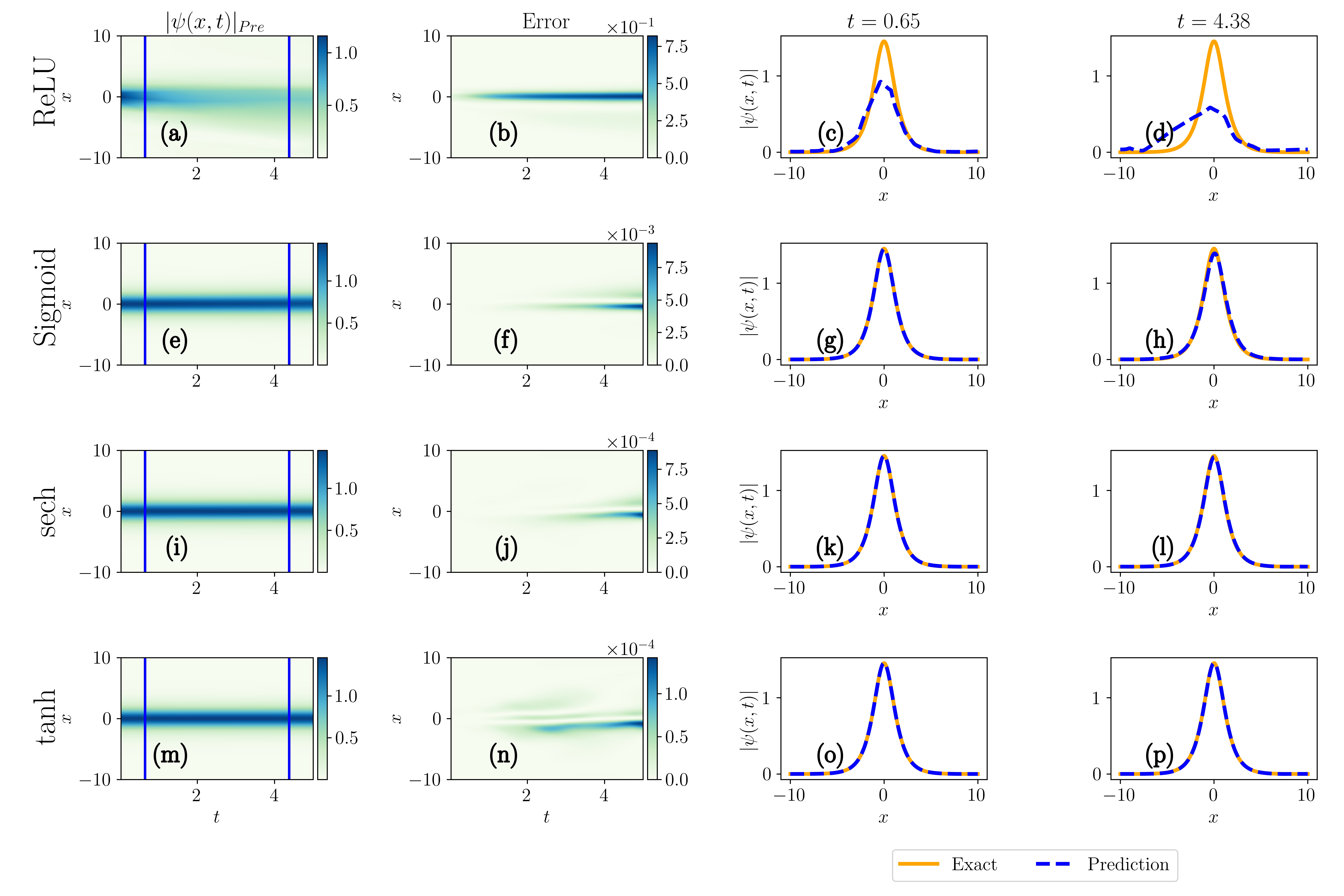}
		\caption{\label{fig:afun_rosen} PINN results for the Rosen-Morse potential with different activation functions. Rows one to four, respectively, represent the activation functions ReLU, sigmoid, sech and tanh. {Figs. (a), (e), (i) and (m) represent the predicted magnitude of soliton solutions. (b), (f), (j) and (n) represent the error values in magnitude of soliton solutions. Figs. (c), (g), (k) and (o) in third column and Figs. (d), (h), (l) and (p) in fourth column correspond to the soliton solution at particular time instants $t=0.65$ and $t=4.38$ respectively}}
	\end{figure*}

	\begin{figure*}[!ht]
		\includegraphics[width=0.75\linewidth]{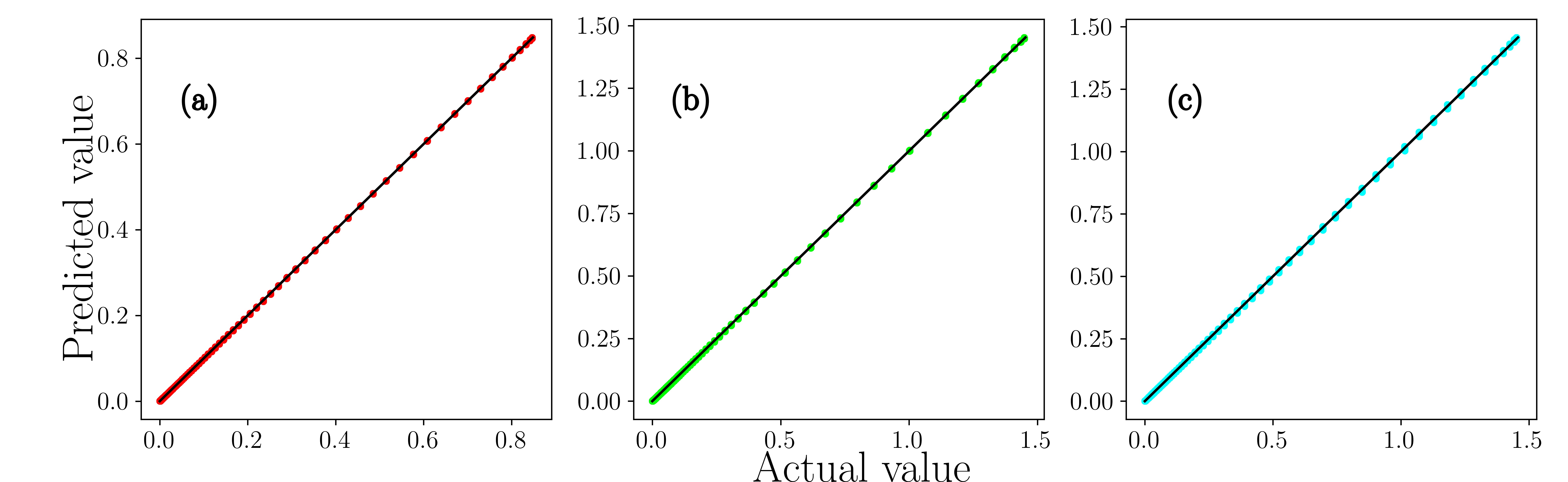}
		\caption{\label{fig:scatt} Scatter plots showing the performance of the PINN. (a)-(c) correspond to the result of NLS equation with Gaussian, periodic and Rosen-Morse respectively. }
	\end{figure*}
	\begin{table*}[!ht]
		\caption{\label{tab:a_fun} $\mathbb{L}^2$-norm error values in $u,v$ and $\psi$ of the PINN results approximated with different activation functions for all three considered potentials.}
		\begin{ruledtabular}
			\begin{tabular}{cccccc}
				& & \multicolumn{4}{c}{Activation functions} \\
				$\mathcal{PT}$-symmetric \\Potentials & $\mathbb{L}^2$-norm Error& ReLU & sigmoid & $\sech$ & $\tanh$\\
				\hline\hline
				& u &  $8.2281\times 10^{-1}$ &  $1.6449\times 10^{-2}$  &  $1.6841\times 10^{-2}$  & $2.1856\times 10^{-2}$   \\
				Gaussian & v &  $7.2510\times 10^{-1}$ & $2.0385\times 10^{-2}$  &   $1.8659\times 10^{-2}$ &  $2.8822\times 10^{-2}$ \\
				& $\psi$ & $6.4185\times 10^{-1}$   &  $1.1915\times 10^{-2}$  &  $5.8949\times 10^{-3}$ &$3.1912\times 10^{-3}$\\
				\hline
				
				& u &   $1.0433\times 10^{0}$ &  $4.7092\times 10^{-2}$  &  $2.9961\times 10^{-2}$  &  $5.0925\times 10^{-2}$  \\
				Periodic & v & $1.0236\times 10^{0}$  & $4.7094\times 10^{-2}$  & $2.9019\times 10^{-2}$   & $4.9897\times 10^{-2}$  \\
				& $\psi$ & $6.9093\times 10^{-1}$   &  $1.3379\times 10^{-2}$  & $5.2880\times 10^{-3}$  &   $4.2720\times 10^{-3}$ \\
				\hline
				& u &  $1.1185\times 10^{0}$  &  $1.1149\times 10^{-1}$  &   $2.4371\times 10^{-2}$ &  $3.7277\times 10^{-2}$  \\
				Rosen-Morse & v &  $1.1529\times 10^{0}$ &  $8.2902\times 10^{-2}$ &   $2.4424\times 10^{-2}$ &   $3.2468\times 10^{-2}$\\
				& $\psi$ &  $4.8206\times 10^{-1}$  & $3.3237\times 10^{-2}$   & $9.8817\times 10^{-3}$ & $5.9112\times 10^{-3}$\\

			\end{tabular}
		\end{ruledtabular}
	\end{table*}
\par Prediction results of PINN with different activation functions for the NLS equation with Rosen-Morse potential are shown in Fig.~\ref{fig:afun_rosen}. The results corresponding to PINN with ReLU as the activation function are shown in Figs.~\ref{fig:afun_rosen} (a)-(d). From Fig.~\ref{fig:afun_rosen} (a) we can see that the approximation done by the PINN is not accurate. From the squared error plots shown in the second column in Fig.~\ref{fig:afun_rosen}, we can see that the error value is low in the case of $\tanh$ function. The plots on the third and fourth columns of Fig.~\ref{fig:afun_rosen} confirm that the predicted and exact results are fitted well with each other while we compare $\tanh$ activation function with the other activation functions.
	\par The overall outcome is presented in the Table \ref{tab:a_fun}. For better comparison, we take the $\mathbb{L}^2$-norm error value in approximating $u$, $v$ and $\psi$ for all the three cases and also for four activation functions. $\mathbb{L}^2$-norm error values of the PINN with ReLU as activation function are very high in approximating $u$, $v$ and $\psi$ for all three potentials when compared with the other three activation functions. {This is due to the piecewise linearity of the ReLU function.} In the case of sigmoid activation function, the value of $\mathbb{L}^2$-norm error is of the order of $10^{-2}$. While comparing the outcome of the PINN with $\sech$ and $\tanh$ functions, it is clear that the error value is slightly low in the case of $\tanh$ while approximating the function $\psi$. But for the approximation of real ($u$) and imaginary ($v$) parts of the solution the PINNs with sech function have low error values for all three considered potentials. {Finally, we note that the time taken for the training of the PINN with $\sech$ activation function is higher when compared to the other three activation functions.}
	\subsection{Effect of structure of the network}
	\par {ANNs have large amount of parameters like weight and bias matrices which change randomly to minimize the given loss function during the process of optimization. So the structure of the ANN influences the accuracy of the PINN for the considered task. There are hyper-parameters that describe the structure of the ANN, namely width of the network (number of hidden layers) and depth of the network (number of units in each layer). We test the impact of these hyper-parameters for all three potentials with $\tanh$ as an activation function and present the outcomes in Tables~{\ref{tab:layers}}, {\ref{tab:nodes1}} and {\ref{tab:nodes2}}.} 
		\begin{table*}[!ht]
		\caption{\label{tab:layers} $\mathbb{L}^2$-norm error values in $u,v$ and $\psi$ of the PINN results approximated with different number of hidden layers of ANN for all three considered potentials.}
		\begin{ruledtabular}
			\begin{tabular}{cccccc}
				& & \multicolumn{4}{c}{Number of hidden layers} \\
				$\mathcal{PT}$-symmetric \\Potentials & $\mathbb{L}^2$-norm Error& 1 & 2 & 3 & 4\\
				\hline\hline
				& u &  $4.0425\times 10^{-2}$ &  $2.0164\times 10^{-2}$  &  $2.3489\times 10^{-2}$ & $2.1856\times 10^{-2}$ \\
				Gaussian & v & $6.9994\times 10^{-2}$  & $2.4169\times 10^{-2}$  &  $3.1864\times 10^{-2}$  & $2.8822\times 10^{-2}$ \\
				& $\psi$ & $2.8418\times 10^{-2}$ & $5.5559\times 10^{-3}$ & $3.5918\times 10^{-3}$ & $3.1912\times 10^{-3}$\\
				\hline
				
				& u & $7.9002\times 10^{-1}$  & $3.2177\times 10^{-2}$ & $4.8171\times 10^{-2}$ &  $5.0925\times 10^{-2}$ \\
				Periodic & v & $9.7382\times 10^{-1}$ & $3.1851\times 10^{-2}$ & $4.7245\times 10^{-2}$ & $4.9897\times 10^{-2}$ \\
				& $\psi$ & $1.8630\times 10^{-1}$ & $7.6633\times 10^{-3}$ & $4.0443\times 10^{-3}$ & $4.2720\times 10^{-3}$\\
				\hline
				
				& u & $3.9728\times 10^{-1}$ & $2.1559\times 10^{-2}$ & $3.7352\times 10^{-2}$ &  $3.7277\times 10^{-2}$\\
				Rosen-Morse & v & $3.7073\times 10^{-1}$ & $2.1047\times 10^{-2}$ & $3.2467\times 10^{-2}$ & $3.2468\times 10^{-2}$\\
				& $\psi$ & $8.3272\times 10^{-2}$ & $6.4215\times 10^{-3}$ & $5.8233\times 10^{-3}$ & $5.9112\times 10^{-3}$\\
			\end{tabular}
		\end{ruledtabular}
	\end{table*}
	{Table~{\ref{tab:layers}} corresponds to the $\mathbb{L}^2$-norm error values of the PINNs with number of hidden layers varying from 1 to 4. In this study, we fix the number of units equal to 100. It is clear from this Table that the performance of the PINN with single hidden layer is very low as compared with the other PINNs with more number of hidden layers. Further, when increasing the number of layers, we observe that the performance of PINN for all three potentials getting increased. In other words, the $\mathbb{L}^2$-norm error values decrease. But in the case of periodic and Rosen-Morse potentials the error values of the PINN with 4 hidden layers are slightly high when compare to the error values of the PINN with 3 hidden layers. The difference between these error values is considerably low. We need a model that performs well in finding solution of the NLS equation for all three considered potentials. So in our study we fixed the number of hidden layers equal to 4.} 
		\begin{table*}[!ht]
		\caption{\label{tab:nodes1} $\mathbb{L}^2$-norm error values in $u,v$ and $\psi$ of the PINN results approximated with different number of neurons (10-50) in each hidden layer of ANN for all three considered potentials.}
		\begin{ruledtabular}
			\begin{tabular}{ccccccc}
				& & \multicolumn{5}{c}{Number of neurons in the hidden layers} \\
				$\mathcal{PT}$-symmetric \\Potentials & $\mathbb{L}^2$-norm Error& 10 & 20 & 30 & 40 & 50\\
				\hline\hline
				& u &  $2.1027\times 10^{-2}$ &  $1.9056\times 10^{-2}$  &  $2.1658\times 10^{-2}$  & $2.0601\times 10^{-2}$ & $1.9279\times 10^{-2}$ \\
				Gaussian & v & $2.7296\times 10^{-2}$  &  $2.2187\times 10^{-2}$  &  $2.7912\times 10^{-2}$  & $2.5865\times 10^{-2}$ & $2.3295\times 10^{-2}$\\
				& $\psi$ & $4.4595\times 10^{-3}$ &  $5.7220
				\times 10^{-3}$ & $3.8554\times 10^{-3}$ & $4.1573\times 10^{-3}$ & $4.5314\times 10^{-3}$\\
				\hline
				
				& u & $5.3167\times 10^{-1}$  & $3.7814\times 10^{-2}$ & $4.2063\times 10^{-2}$ &  $4.5505\times 10^{-2}$ & $4.6104\times 10^{-2}$\\
				Periodic & v & $7.0910\times 10^{-1}$ & $3.6862\times 10^{-2}$ & $4.1328\times 10^{-2}$ & $4.4600\times 10^{-2}$ & $4.5129\times 10^{-2}$\\
				& $\psi$ & $1.3475\times 10^{-1}$ & $4.3729\times 10^{-3}$ & $4.1834\times 10^{-3}$ & $4.0019\times 10^{-3}$ & $3.7271\times 10^{-3}$\\
				\hline
				
				& u &  $1.4408\times 10^{-1}$ & $2.5117\times 10^{-2}$ & $3.9776\times 10^{-2}$ &  $2.7235\times 10^{-2}$ & $3.1777\times 10^{-2}$\\
				Rosen-Morse & v &  $1.0567\times 10^{-1}$ & $2.3972\times 10^{-2}$ & $3.3787\times 10^{-2}$ & $2.5332\times 10^{-2}$ & $2.7809\times 10^{-2}$\\
				& $\psi$ &  $3.2571\times 10^{-2}$ & $7.2073\times 10^{-3}$ & $5.2887\times 10^{-3}$ & $5.1823\times 10^{-3}$ & $5.0052\times 10^{-3}$\\

			\end{tabular}
		\end{ruledtabular}
	\end{table*}
			\begin{table*}[!ht]
		\caption{\label{tab:nodes2} $\mathbb{L}^2$-norm error values in $u,v$ and $\psi$ of the PINN results approximated with different number of neurons (60-100) in each hidden layer of ANN for all three considered potentials.}
		\begin{ruledtabular}
			\begin{tabular}{ccccccc}
				& & \multicolumn{5}{c}{Number of neurons in the hidden layers} \\
				$\mathcal{PT}$-symmetric \\Potentials & $\mathbb{L}^2$-norm Error& 60 & 70 & 80 & 90 & 100\\
				\hline\hline
				& u & $1.8949\times 10^{-2}$ & $1.9171\times 10^{-2}$ & $1.8510\times 10^{-2}$ & $2.0515\times 10^{-2}$ & $2.1856\times 10^{-2}$ \\
				Gaussian & v & $2.2847\times 10^{-2}$ & $2.3018\times 10^{-2}$ & $2.1814\times 10^{-2}$ &  $2.5669\times 10^{-2}$ & $2.8822\times 10^{-2}$ \\
				& $\psi$ & $4.3099\times 10^{-3}$ & $4.6188\times 10^{-3}$ & $4.9689\times 10^{-3}$ & $4.0189\times 10^{-3}$ & $3.1912\times 10^{-3}$\\
				\hline
				
				& u & $4.8924\times 10^{-2}$  & $4.3401\times 10^{-2}$ & $5.2175\times 10^{-2}$ &  $5.0627\times 10^{-2}$ & $5.0925\times 10^{-2}$\\
				Periodic & v & $4.7821\times 10^{-2}$ & $4.2629\times 10^{-2}$ & $5.0985\times 10^{-2}$ & $4.9554\times 10^{-2}$ & $4.9897\times 10^{-2}$ \\
				& $\psi$ & $3.8299\times 10^{-3}$ & $3.8624\times 10^{-3}$ & $4.0359\times 10^{-3}$ & $3.9781\times 10^{-3}$ & $4.2720\times 10^{-3}$\\
				\hline
				
				& u & $3.0926\times 10^{-2}$ & $3.7381\times 10^{-2}$ & $4.4708\times 10^{-2}$ & $4.1246\times 10^{-2}$ &  $3.7277\times 10^{-2}$\\
				Rosen-Morse & v & $2.7566\times 10^{-2}$ & $3.2622\times 10^{-2}$ & $3.7069\times 10^{-2}$ & $3.5090\times 10^{-2}$ & $3.2468\times 10^{-2}$\\
				& $\psi$ & $4.9355\times 10^{-3}$ & $5.8375\times 10^{-3}$ & $5.3917\times 10^{-3}$ & $5.6393\times 10^{-3}$ & $5.9112\times 10^{-3}$\\

			\end{tabular}
		\end{ruledtabular}
	\end{table*}
	\par {Next we examine how the performance of PINN is affected by the number of neurons in the hidden layers. For this we consider the PINN with $\tanh$ activation function and four hidden layers. Now we vary the the number of neurons from 10 to 100 and present the results of the PINN with the number of neurons 10-50 in Table~{\ref{tab:nodes1}} and for the number of neurons 60-100 in Table~{\ref{tab:nodes2}} respectively. From these two tables, we can see that $\mathbb{L}^2$-norm error values are very high for the case of PINN with ten neurons. Further, increasing the number of neurons the error values are decreasing and for some cases they are oscillating between low and high values because while increasing the number of neurons automatically increases the size of the weight and bias matrices and the model needs to optimize the more number of parameters. The error value of the solution of the NLS equation with the Gaussian potential gives a low value only when the PINN is trained with 100 neurons. So we fixed the number of neurons in each hidden layer as 100.}
	 
		\begin{table*}[!ht]
		\caption{\label{tab:c_points} $\mathbb{L}^2$-norm error values in $u,v$ and $\psi$ of the PINN results approximated with different number of collocation points for all three considered potentials.}
		\begin{ruledtabular}
			\begin{tabular}{cccccc}
				& & \multicolumn{4}{c}{Number of collocation points} \\
				$\mathcal{PT}$-symmetric \\Potentials & $\mathbb{L}^2$-norm Error& 5000 & 10000 & 15000 & 20000\\
				\hline\hline
				& u & $2.0919\times 10^{-2}$ & $1.9999\times 10^{-2}$ & $1.9238\times 10^{-2}$ & $2.1856\times 10^{-2}$ \\
				Gaussian & v & $2.6675\times 10^{-2}$ & $2.4926\times 10^{-2}$ & $2.3326\times 10^{-2}$ & $2.8822\times 10^{-2}$ \\
				& $\psi$ & $3.6122\times 10^{-3}$ & $4.0438\times 10^{-3}$ & $4.1420\times 10^{-3}$ & $3.1912\times 10^{-3}$\\
				\hline
				
				& u & $4.5870\times 10^{-2}$  & $5.1518\times 10^{-2}$ & $4.2062\times 10^{-2}$ &  $5.0925\times 10^{-2}$ \\
				Periodic & v & $4.4983\times 10^{-2}$ & $5.0452\times 10^{-2}$  & $4.1056\times 10^{-2}$ & $4.9897\times 10^{-2}$ \\
				& $\psi$ & $3.8983\times 10^{-3}$ & $3.8867\times 10^{-3}$ & $3.8140\times 10^{-3}$ & $4.2720\times 10^{-3}$\\
				\hline
				
				& u & $3.8685\times 10^{-2}$ & $3.7610\times 10^{-2}$ & $4.6277\times 10^{-2}$ &  $3.7277\times 10^{-2}$\\
				Rosen-Morse & v & $3.3381\times 10^{-2}$ & $3.2416\times 10^{-2}$ & $3.8593\times 10^{-2}$ & $3.2468\times 10^{-2}$\\
				& $\psi$ & $6.0374\times 10^{-3}$ & $5.6554\times 10^{-3}$ & $5.8951\times 10^{-3}$ & $5.9112\times 10^{-3}$\\

			\end{tabular}
		\end{ruledtabular}
	\end{table*}

		\begin{table*}[!ht]
		\caption{\label{tab:ini_bound} $\mathbb{L}^2$-norm error values in $u,v$ and $\psi$ of the PINN results approximated with different number of  initial and boundary points of the considered domain for all three considered potentials.}
		\begin{ruledtabular}
			\begin{tabular}{ccccccc}
				& & \multicolumn{5}{c}{Number of initial and boundary points} \\
				$\mathcal{PT}$-symmetric \\Potentials & $\mathbb{L}^2$-norm Error& 10 & 20 & 30 & 40 & 50\\
				\hline\hline
				& u & $1.7453\times 10^{-2}$ & $1.7899\times 10^{-2}$ & $1.6184\times 10^{-2}$ & $2.1619\times 10^{-2}$ & $2.1856\times 10^{-2}$ \\
				Gaussian & v & $2.0003\times 10^{-2}$ & $2.0321\times 10^{-2}$ & $1.7222\times 10^{-2}$ & $2.8040\times 10^{-2}$ & $2.8822\times 10^{-2}$ \\
				& $\psi$ & $5.5589\times 10^{-3}$ & $4.9169\times 10^{-3}$ & $6.3697\times 10^{-3}$ & $3.6132\times 10^{-3}$ & $3.1912\times 10^{-3}$\\
				\hline
				
				& u & $2.5785\times 10^{-2}$ & $4.5283\times 10^{-2}$  & $3.3628\times 10^{-2}$ & $4.9036\times 10^{-2}$ &  $5.0925\times 10^{-2}$ \\
				Periodic & v & $2.4970\times 10^{-2}$ & $4.4399\times 10^{-2}$ & $3.2921\times 10^{-2}$ & $4.8044\times 10^{-2}$ & $4.9897\times 10^{-2}$ \\
				& $\psi$ & $6.9655\times 10^{-3}$ & $4.0064\times 10^{-3}$ & $4.3703\times 10^{-3}$ & $3.9586\times 10^{-3}$ & $4.2720\times 10^{-3}$\\
				\hline
				
				& u & $2.1818\times 10^{-2}$ & $3.6917\times 10^{-2}$ & $3.4343\times 10^{-2}$ & $3.9939\times 10^{-2}$ &  $3.7277\times 10^{-2}$\\
				Rosen-Morse & v & $2.2776\times 10^{-2}$ & $3.1811\times 10^{-2}$ & $2.9756\times 10^{-2}$ & $3.4444\times 10^{-2}$ & $3.2468\times 10^{-2}$\\
				& $\psi$ & $1.0650\times 10^{-2}$ & $5.2218\times 10^{-3}$ & $4.9235\times 10^{-3}$ & $5.0712\times 10^{-3}$ & $5.9112\times 10^{-3}$\\

			\end{tabular}
		\end{ruledtabular}
	\end{table*}
	\subsection{Effect of sampling points}
	\par {We use the sampling points which are sampled from LHS~{\cite{stein1987}} for the input to the PINN model. These training data points also influence the performance of the PINN in solving the considered problem. The results with different number of collocation points are presented in Table~{\ref{tab:c_points}}. For this study we use $\tanh$ activation function, four hidden layers each with 100 neurons and 50 initial points and 50 points each on upper and lower boundaries. From Table~{\ref{tab:c_points}} it is clear that $\mathbb{L}^2$-norm error values are changing with respect to the change in the number of collocation points. When the number of collocation points is 20000 the error value is completely low for all cases especially for the Gaussian potential case. Our aim is to construct a DL model which is good enough to make the solution to the NLS equation for all three considered potentials. So it is better to have a more number of collocation points inside the considered domain so that the model can train with more points which lead to high accurate solution.} 
\par {Finally, we experiment the PINN by varying the number of initial and boundary sampling points and the $\mathbb{L}^2$-norm error values in $u,v$ and $\psi$ for all three potentials presented in Table~{\ref{tab:ini_bound}}. In this table, we vary the number of points from 10 to 50 in both the initial and boundary regions. Here the number on the boundary denotes the number of points taken for both the upper and lower boundary. For example, the number 10 denotes that there are 10 points on the initial and also 10 points each on both the upper and lower boundary regions so that there are totally 20 points on the boundary of the considered domain. Here also we use PINN with $\tanh$ activation function with 4 hidden layers each with 100 neurons and 20000 collocation points inside the domain. From the results which are shown in Table~{\ref{tab:ini_bound}} we observe that in most cases the error values are become low when we increase the number of points and also in some cases the error values vary between low and high values particularly in the case of NLS equation with Rosen-Morse potential the error value is very high when the PINN trained with low number of points on the initial and boundary regions say 10. As discussed in the earlier cases, here also we fix the number of points on the initial and boundary is equal to 50 because the PINN with this setup has considerable low values for the $\mathbb{L}^2$-norm error for the all three considered potentials.}
	\par {It is worth to note that all the above presented results may vary in the repeated learning processes because of the stochastic nature of the sampling technique and of the algorithm.}
	
	\section{Conclusion}
	\par In this work, we have considered the NLS equation with three $\mathcal{PT}$-symmetric potentials, namely Gaussian, periodic and Rosen-Morse and approximated the soliton solution of the NLS equation with the help of a DL approach so called PINN. For this purpose, we have considered a complex-valued PINN with $\tanh$ as an activation function. The PINN solves the given equation for the prescribed initial and boundary conditions by minimizing the mean squared error loss. We have considered 20000 collocation points by LHS \cite{stein1987}, 50 points and 100 points on initial and boundary data respectively. The predicted, exact and squared error in the magnitude of soliton solution for the considered three different potentials are evaluated and plotted. Further, we have also plotted the exact and predicted magnitudes of the soliton solution one over the other for various instants of time. From the results, we conclude that our constructed PINN can approximate the soliton solution for the given NLS equation for all three potentials precisely. The squared errors are found to be very low in the order of $10^{-6}$, $10^{-5}$ and $10^{-4}$ respectively for the Gaussian, periodic and Rosen-Morse potentials. 
	
	\par To visualize the performance of the PINN with tanh as activation function, we also present the scatter plot of actual versus the predicted data for all three considered potentials in Fig.~\ref{fig:scatt}. The scatter plots of the NLS equation with Gaussian, periodic and Rosen-Morse potentials are respectively shown in Figs.~\ref{fig:scatt} (a)-(c). The scatter plots confirm that the considered PINN accurately predict the soliton solution in all three cases. {Further, to analyse the factors that influence the performance of the PINN we tested the effect against the activation functions, network structure and sampling points. First, We have considered three functions namely, ReLU, sigmoid and $\tanh$ along with a new activation function $\sech$. The PINNs with ReLU and sigmoid as the activation functions approximated the soliton solution with less accuracy when compared to the PINNs with $\sech$ and $\tanh$ as activation functions. We have also examined the ability of these different PINNs by calculating the $\mathbb{L}^2$-norm error values for real ($u$) and imaginary ($v$) parts of the solution ($\psi$) for all three considered potentials. From the results, we conclude that the PINN can approximate the soliton solution of the NLS equation for the considered $\mathcal{PT}$-symmetric potentials with $\tanh$ and $\sech$ as activation function. We have also examined the effect on the performance due to the width and the depth of the PINN. From the obtained results we fixed the number of hidden layers equal to four and 100 neurons in each layer. Finally, we have also done an experiment on the number of sampling points and initial and boundary regions. From the outcomes we found that the amount of training data should be 20000 collocation points, 50 initial points and 50 boundary points in order to get a high accurate solution for the considered problem. One can use the considered DL model, namely PINN for solving the NLS equation with $\mathcal{PT}$-symmetric potentials.}

	\begin{acknowledgments}
		JM thanks MoE - RUSA 2.0 Physical Sciences, Government of lndia for providing a fellowship to carry out this work. KM and JBS are funded by the Center for Nonlinear Systems, Chennai Institute of Technology, India, vide funding number CIT/CNS/2021/RP-015. The work of MS forms part of a research project sponsored by NBHM, Government of India, under the Grant No. 02011/20/2018 NBHM (R.P)/R\&D II/15064. MS also acknowledges MoE - RUSA 2.0 Physical Sciences, Government of lndia for providing financial support in procuring a high-performance GPU server which highly assisted this work.
	\end{acknowledgments}
	
	\section*{Data Availability Statement}
	
	The data that support the findings of this study are available within the article.

	\appendix

	\bibliography{references}

\providecommand{\noopsort}[1]{}\providecommand{\singleletter}[1]{#1}%
\begin{thebibliography}{49}%
\makeatletter
\providecommand \@ifxundefined [1]{%
 \@ifx{#1\undefined}
}%
\providecommand \@ifnum [1]{%
 \ifnum #1\expandafter \@firstoftwo
 \else \expandafter \@secondoftwo
 \fi
}%
\providecommand \@ifx [1]{%
 \ifx #1\expandafter \@firstoftwo
 \else \expandafter \@secondoftwo
 \fi
}%
\providecommand \natexlab [1]{#1}%
\providecommand \enquote  [1]{``#1''}%
\providecommand \bibnamefont  [1]{#1}%
\providecommand \bibfnamefont [1]{#1}%
\providecommand \citenamefont [1]{#1}%
\providecommand \href@noop [0]{\@secondoftwo}%
\providecommand \href [0]{\begingroup \@sanitize@url \@href}%
\providecommand \@href[1]{\@@startlink{#1}\@@href}%
\providecommand \@@href[1]{\endgroup#1\@@endlink}%
\providecommand \@sanitize@url [0]{\catcode `\\12\catcode `\$12\catcode
  `\&12\catcode `\#12\catcode `\^12\catcode `\_12\catcode `\%12\relax}%
\providecommand \@@startlink[1]{}%
\providecommand \@@endlink[0]{}%
\providecommand \url  [0]{\begingroup\@sanitize@url \@url }%
\providecommand \@url [1]{\endgroup\@href {#1}{\urlprefix }}%
\providecommand \urlprefix  [0]{URL }%
\providecommand \Eprint [0]{\href }%
\providecommand \doibase [0]{http://dx.doi.org/}%
\providecommand \selectlanguage [0]{\@gobble}%
\providecommand \bibinfo  [0]{\@secondoftwo}%
\providecommand \bibfield  [0]{\@secondoftwo}%
\providecommand \translation [1]{[#1]}%
\providecommand \BibitemOpen [0]{}%
\providecommand \bibitemStop [0]{}%
\providecommand \bibitemNoStop [0]{.\EOS\space}%
\providecommand \EOS [0]{\spacefactor3000\relax}%
\providecommand \BibitemShut  [1]{\csname bibitem#1\endcsname}%
\let\auto@bib@innerbib\@empty
\bibitem [{\citenamefont {Malomed}\ and\ \citenamefont
  {Mihalache}(2019)}]{malomed}%
  \BibitemOpen
  \bibfield  {author} {\bibinfo {author} {\bibfnamefont {B.~A.}\ \bibnamefont
  {Malomed}}\ and\ \bibinfo {author} {\bibfnamefont {D.}~\bibnamefont
  {Mihalache}},\ }\bibfield  {title} {\enquote {\bibinfo {title} {Nonlinear
  waves in optical and matter-wave media: A topical survey of recent
  theoretical and experimental results},}\ }\href@noop {} {\bibfield  {journal}
  {\bibinfo  {journal} {Rom. J. Phys.}\ }\textbf {\bibinfo {volume} {64}},\
  \bibinfo {pages} {106} (\bibinfo {year} {2019})}\BibitemShut {NoStop}%
\bibitem [{\citenamefont {Bender}\ and\ \citenamefont
  {Boettcher}(1998)}]{bender1998}%
  \BibitemOpen
  \bibfield  {author} {\bibinfo {author} {\bibfnamefont {C.~M.}\ \bibnamefont
  {Bender}}\ and\ \bibinfo {author} {\bibfnamefont {S.}~\bibnamefont
  {Boettcher}},\ }\bibfield  {title} {\enquote {\bibinfo {title} {Real spectra
  in non-hermitian hamiltonians having $\mathcal{PT}$ symmetry},}\ }\href@noop
  {} {\bibfield  {journal} {\bibinfo  {journal} {Phys. Rev. Lett.}\ }\textbf
  {\bibinfo {volume} {80}},\ \bibinfo {pages} {5243} (\bibinfo {year}
  {1998})}\BibitemShut {NoStop}%
\bibitem [{\citenamefont {Musslimani}\ \emph {et~al.}(2008)\citenamefont
  {Musslimani}, \citenamefont {Makris}, \citenamefont {El-Ganainy},\ and\
  \citenamefont {Christodoulides}}]{musslimani2008optical}%
  \BibitemOpen
  \bibfield  {author} {\bibinfo {author} {\bibfnamefont {Z.}~\bibnamefont
  {Musslimani}}, \bibinfo {author} {\bibfnamefont {K.~G.}\ \bibnamefont
  {Makris}}, \bibinfo {author} {\bibfnamefont {R.}~\bibnamefont {El-Ganainy}},
  \ and\ \bibinfo {author} {\bibfnamefont {D.~N.}\ \bibnamefont
  {Christodoulides}},\ }\bibfield  {title} {\enquote {\bibinfo {title} {Optical
  solitons in $\mathcal{PT}$ periodic potentials},}\ }\href@noop {} {\bibfield
  {journal} {\bibinfo  {journal} {Phys. Rev. Lett.}\ }\textbf {\bibinfo
  {volume} {100}},\ \bibinfo {pages} {030402} (\bibinfo {year}
  {2008})}\BibitemShut {NoStop}%
\bibitem [{\citenamefont {Kominis}\ \emph {et~al.}(2019)\citenamefont
  {Kominis}, \citenamefont {Cuevas-Maraver}, \citenamefont {Kevrekidis},
  \citenamefont {Frantzeskakis},\ and\ \citenamefont
  {Bountis}}]{kominis2019continuous}%
  \BibitemOpen
  \bibfield  {author} {\bibinfo {author} {\bibfnamefont {Y.}~\bibnamefont
  {Kominis}}, \bibinfo {author} {\bibfnamefont {J.}~\bibnamefont
  {Cuevas-Maraver}}, \bibinfo {author} {\bibfnamefont {P.~G.}\ \bibnamefont
  {Kevrekidis}}, \bibinfo {author} {\bibfnamefont {D.~J.}\ \bibnamefont
  {Frantzeskakis}}, \ and\ \bibinfo {author} {\bibfnamefont {A.}~\bibnamefont
  {Bountis}},\ }\bibfield  {title} {\enquote {\bibinfo {title} {Continuous
  families of solitary waves in non-symmetric complex potentials: A melnikov
  theory approach},}\ }\href@noop {} {\bibfield  {journal} {\bibinfo  {journal}
  {Chaos, Solitons \& Fractals}\ }\textbf {\bibinfo {volume} {118}},\ \bibinfo
  {pages} {222--233} (\bibinfo {year} {2019})}\BibitemShut {NoStop}%
\bibitem [{\citenamefont {Yan}, \citenamefont {Wen},\ and\ \citenamefont
  {Hang}(2015)}]{zyan}%
  \BibitemOpen
  \bibfield  {author} {\bibinfo {author} {\bibfnamefont {Z.}~\bibnamefont
  {Yan}}, \bibinfo {author} {\bibfnamefont {Z.}~\bibnamefont {Wen}}, \ and\
  \bibinfo {author} {\bibfnamefont {C.}~\bibnamefont {Hang}},\ }\bibfield
  {title} {\enquote {\bibinfo {title} {Spatial solitons and stability in
  self-focusing and defocusing kerr nonlinear media with generalized
  parity-time-symmetric {Scarff-II} potentials},}\ }\href@noop {} {\bibfield
  {journal} {\bibinfo  {journal} {Phys. Rev. E}\ }\textbf {\bibinfo {volume}
  {92}},\ \bibinfo {pages} {022913} (\bibinfo {year} {2015})}\BibitemShut
  {NoStop}%
\bibitem [{\citenamefont {Guo}\ \emph {et~al.}(2009)\citenamefont {Guo},
  \citenamefont {Salamo}, \citenamefont {Duchesne}, \citenamefont {Morandotti},
  \citenamefont {Volatier-Ravat}, \citenamefont {Aimez}, \citenamefont
  {Siviloglou},\ and\ \citenamefont {Christodoulides}}]{guo2009observation}%
  \BibitemOpen
  \bibfield  {author} {\bibinfo {author} {\bibfnamefont {A.}~\bibnamefont
  {Guo}}, \bibinfo {author} {\bibfnamefont {G.}~\bibnamefont {Salamo}},
  \bibinfo {author} {\bibfnamefont {D.}~\bibnamefont {Duchesne}}, \bibinfo
  {author} {\bibfnamefont {R.}~\bibnamefont {Morandotti}}, \bibinfo {author}
  {\bibfnamefont {M.}~\bibnamefont {Volatier-Ravat}}, \bibinfo {author}
  {\bibfnamefont {V.}~\bibnamefont {Aimez}}, \bibinfo {author} {\bibfnamefont
  {G.}~\bibnamefont {Siviloglou}}, \ and\ \bibinfo {author} {\bibfnamefont
  {D.}~\bibnamefont {Christodoulides}},\ }\bibfield  {title} {\enquote
  {\bibinfo {title} {Observation of $\mathcal{PT}$-symmetry breaking in complex
  optical potentials},}\ }\href@noop {} {\bibfield  {journal} {\bibinfo
  {journal} {Phys. Rev. Lett.}\ }\textbf {\bibinfo {volume} {103}},\ \bibinfo
  {pages} {093902} (\bibinfo {year} {2009})}\BibitemShut {NoStop}%
\bibitem [{\citenamefont {R{\"u}ter}\ \emph {et~al.}(2010)\citenamefont
  {R{\"u}ter}, \citenamefont {Makris}, \citenamefont {El-Ganainy},
  \citenamefont {Christodoulides}, \citenamefont {Segev},\ and\ \citenamefont
  {Kip}}]{Ruter}%
  \BibitemOpen
  \bibfield  {author} {\bibinfo {author} {\bibfnamefont {C.~E.}\ \bibnamefont
  {R{\"u}ter}}, \bibinfo {author} {\bibfnamefont {K.~G.}\ \bibnamefont
  {Makris}}, \bibinfo {author} {\bibfnamefont {R.}~\bibnamefont {El-Ganainy}},
  \bibinfo {author} {\bibfnamefont {D.~N.}\ \bibnamefont {Christodoulides}},
  \bibinfo {author} {\bibfnamefont {M.}~\bibnamefont {Segev}}, \ and\ \bibinfo
  {author} {\bibfnamefont {D.}~\bibnamefont {Kip}},\ }\bibfield  {title}
  {\enquote {\bibinfo {title} {Observation of parity-time symmetry in
  optics},}\ }\href@noop {} {\bibfield  {journal} {\bibinfo  {journal} {Nat.
  Phys.}\ }\textbf {\bibinfo {volume} {6}},\ \bibinfo {pages} {192--195}
  (\bibinfo {year} {2010})}\BibitemShut {NoStop}%
\bibitem [{\citenamefont {Regensburger}\ \emph {et~al.}(2013)\citenamefont
  {Regensburger}, \citenamefont {Miri}, \citenamefont {Bersch}, \citenamefont
  {N{\"a}ger}, \citenamefont {Onishchukov}, \citenamefont {Christodoulides},\
  and\ \citenamefont {Peschel}}]{regensburger}%
  \BibitemOpen
  \bibfield  {author} {\bibinfo {author} {\bibfnamefont {A.}~\bibnamefont
  {Regensburger}}, \bibinfo {author} {\bibfnamefont {M.}~\bibnamefont {Miri}},
  \bibinfo {author} {\bibfnamefont {C.}~\bibnamefont {Bersch}}, \bibinfo
  {author} {\bibfnamefont {J.}~\bibnamefont {N{\"a}ger}}, \bibinfo {author}
  {\bibfnamefont {G.}~\bibnamefont {Onishchukov}}, \bibinfo {author}
  {\bibfnamefont {D.~N.}\ \bibnamefont {Christodoulides}}, \ and\ \bibinfo
  {author} {\bibfnamefont {U.}~\bibnamefont {Peschel}},\ }\bibfield  {title}
  {\enquote {\bibinfo {title} {Observation of defect states in
  $\mathcal{PT}$-symmetric optical lattices},}\ }\href@noop {} {\bibfield
  {journal} {\bibinfo  {journal} {Phys. Rev. Lett.}\ }\textbf {\bibinfo
  {volume} {110}},\ \bibinfo {pages} {223902} (\bibinfo {year}
  {2013})}\BibitemShut {NoStop}%
\bibitem [{\citenamefont {Makris}\ \emph {et~al.}(2008)\citenamefont {Makris},
  \citenamefont {El-Ganainy}, \citenamefont {Christodoulides},\ and\
  \citenamefont {Musslimani}}]{Makris}%
  \BibitemOpen
  \bibfield  {author} {\bibinfo {author} {\bibfnamefont {K.~G.}\ \bibnamefont
  {Makris}}, \bibinfo {author} {\bibfnamefont {R.}~\bibnamefont {El-Ganainy}},
  \bibinfo {author} {\bibfnamefont {D.}~\bibnamefont {Christodoulides}}, \ and\
  \bibinfo {author} {\bibfnamefont {Z.~H.}\ \bibnamefont {Musslimani}},\
  }\bibfield  {title} {\enquote {\bibinfo {title} {Beam dynamics in
  $\mathcal{PT}$-symmetric optical lattices},}\ }\href@noop {} {\bibfield
  {journal} {\bibinfo  {journal} {Phys. Rev. Lett.}\ }\textbf {\bibinfo
  {volume} {100}},\ \bibinfo {pages} {103904} (\bibinfo {year}
  {2008})}\BibitemShut {NoStop}%
\bibitem [{\citenamefont {Chen}\ \emph {et~al.}(2017)\citenamefont {Chen},
  \citenamefont {Yan}, \citenamefont {Mihalache},\ and\ \citenamefont
  {Malomed}}]{Chen2}%
  \BibitemOpen
  \bibfield  {author} {\bibinfo {author} {\bibfnamefont {Y.}~\bibnamefont
  {Chen}}, \bibinfo {author} {\bibfnamefont {Z.}~\bibnamefont {Yan}}, \bibinfo
  {author} {\bibfnamefont {D.}~\bibnamefont {Mihalache}}, \ and\ \bibinfo
  {author} {\bibfnamefont {B.~A.}\ \bibnamefont {Malomed}},\ }\bibfield
  {title} {\enquote {\bibinfo {title} {Families of stable solitons and
  excitations in the {$\mathcal{PT}$}-symmetric nonlinear {Schr{\"o}dinger}
  equations with position-dependent effective masses},}\ }\href@noop {}
  {\bibfield  {journal} {\bibinfo  {journal} {Scientific Reports}\ }\textbf
  {\bibinfo {volume} {7}},\ \bibinfo {pages} {1--21} (\bibinfo {year}
  {2017})}\BibitemShut {NoStop}%
\bibitem [{\citenamefont {Wen}\ and\ \citenamefont {Yan}(2017)}]{wen}%
  \BibitemOpen
  \bibfield  {author} {\bibinfo {author} {\bibfnamefont {Z.}~\bibnamefont
  {Wen}}\ and\ \bibinfo {author} {\bibfnamefont {Z.}~\bibnamefont {Yan}},\
  }\bibfield  {title} {\enquote {\bibinfo {title} {Solitons and their stability
  in the nonlocal nonlinear {Schr{\"o}dinger} equation with
  $\mathcal{PT}$-symmetric potentials},}\ }\href@noop {} {\bibfield  {journal}
  {\bibinfo  {journal} {Chaos}\ }\textbf {\bibinfo {volume} {27}},\ \bibinfo
  {pages} {053105} (\bibinfo {year} {2017})}\BibitemShut {NoStop}%
\bibitem [{\citenamefont {Hari}, \citenamefont {Manikandan},\ and\
  \citenamefont {Sankaranarayanan}(2020)}]{hari}%
  \BibitemOpen
  \bibfield  {author} {\bibinfo {author} {\bibfnamefont {K.}~\bibnamefont
  {Hari}}, \bibinfo {author} {\bibfnamefont {K.}~\bibnamefont {Manikandan}}, \
  and\ \bibinfo {author} {\bibfnamefont {R.}~\bibnamefont {Sankaranarayanan}},\
  }\bibfield  {title} {\enquote {\bibinfo {title} {Dissipative optical solitons
  in asymmetric {Rosen-Morse} potential},}\ }\href@noop {} {\bibfield
  {journal} {\bibinfo  {journal} {Phys. Lett. A}\ }\textbf {\bibinfo {volume}
  {384}},\ \bibinfo {pages} {126104} (\bibinfo {year} {2020})}\BibitemShut
  {NoStop}%
\bibitem [{\citenamefont {Manikandan}\ \emph {et~al.}(2018)\citenamefont
  {Manikandan}, \citenamefont {Vishnu~Priya}, \citenamefont {Senthilvelan},\
  and\ \citenamefont {Sankaranarayanan}}]{manikandan2018deformation}%
  \BibitemOpen
  \bibfield  {author} {\bibinfo {author} {\bibfnamefont {K.}~\bibnamefont
  {Manikandan}}, \bibinfo {author} {\bibfnamefont {N.}~\bibnamefont
  {Vishnu~Priya}}, \bibinfo {author} {\bibfnamefont {M.}~\bibnamefont
  {Senthilvelan}}, \ and\ \bibinfo {author} {\bibfnamefont {R.}~\bibnamefont
  {Sankaranarayanan}},\ }\bibfield  {title} {\enquote {\bibinfo {title}
  {Deformation of dark solitons in a $\mathcal{PT}$-invariant variable
  coefficients nonlocal nonlinear {Schr{\"o}dinger} equation},}\ }\href@noop {}
  {\bibfield  {journal} {\bibinfo  {journal} {Chaos}\ }\textbf {\bibinfo
  {volume} {28}},\ \bibinfo {pages} {083103} (\bibinfo {year}
  {2018})}\BibitemShut {NoStop}%
\bibitem [{\citenamefont {Manikandan}, \citenamefont {Sudharsan},\ and\
  \citenamefont {Senthilvelan}(2021)}]{manikandan2021nonlinear}%
  \BibitemOpen
  \bibfield  {author} {\bibinfo {author} {\bibfnamefont {K.}~\bibnamefont
  {Manikandan}}, \bibinfo {author} {\bibfnamefont {J.}~\bibnamefont
  {Sudharsan}}, \ and\ \bibinfo {author} {\bibfnamefont {M.}~\bibnamefont
  {Senthilvelan}},\ }\bibfield  {title} {\enquote {\bibinfo {title} {Nonlinear
  tunneling of solitons in a variable coefficients nonlinear {Schr{\"o}dinger}
  equation with $\mathcal{PT}$-symmetric {Rosen-Morse} potential},}\
  }\href@noop {} {\bibfield  {journal} {\bibinfo  {journal} {Eur. Phys. J. B}\
  }\textbf {\bibinfo {volume} {94}},\ \bibinfo {pages} {1--10} (\bibinfo {year}
  {2021})}\BibitemShut {NoStop}%
\bibitem [{\citenamefont {Carleo}\ \emph {et~al.}(2019)\citenamefont {Carleo},
  \citenamefont {Cirac}, \citenamefont {Cranmer}, \citenamefont {Daudet},
  \citenamefont {Schuld}, \citenamefont {Tishby}, \citenamefont
  {Vogt-Maranto},\ and\ \citenamefont {Zdeborov\'a}}]{Carleo2019}%
  \BibitemOpen
  \bibfield  {author} {\bibinfo {author} {\bibfnamefont {G.}~\bibnamefont
  {Carleo}}, \bibinfo {author} {\bibfnamefont {I.}~\bibnamefont {Cirac}},
  \bibinfo {author} {\bibfnamefont {K.}~\bibnamefont {Cranmer}}, \bibinfo
  {author} {\bibfnamefont {L.}~\bibnamefont {Daudet}}, \bibinfo {author}
  {\bibfnamefont {M.}~\bibnamefont {Schuld}}, \bibinfo {author} {\bibfnamefont
  {N.}~\bibnamefont {Tishby}}, \bibinfo {author} {\bibfnamefont
  {L.}~\bibnamefont {Vogt-Maranto}}, \ and\ \bibinfo {author} {\bibfnamefont
  {L.}~\bibnamefont {Zdeborov\'a}},\ }\bibfield  {title} {\enquote {\bibinfo
  {title} {Machine learning and the physical sciences},}\ }\href@noop {}
  {\bibfield  {journal} {\bibinfo  {journal} {Rev. Mod. Phys.}\ }\textbf
  {\bibinfo {volume} {91}},\ \bibinfo {pages} {045002} (\bibinfo {year}
  {2019})}\BibitemShut {NoStop}%
\bibitem [{\citenamefont {Choudhary}\ \emph {et~al.}(2020)\citenamefont
  {Choudhary}, \citenamefont {Lindner}, \citenamefont {Holliday}, \citenamefont
  {Miller}, \citenamefont {Sinha},\ and\ \citenamefont {Ditto}}]{sudhe1}%
  \BibitemOpen
  \bibfield  {author} {\bibinfo {author} {\bibfnamefont {A.}~\bibnamefont
  {Choudhary}}, \bibinfo {author} {\bibfnamefont {J.~F.}\ \bibnamefont
  {Lindner}}, \bibinfo {author} {\bibfnamefont {E.~G.}\ \bibnamefont
  {Holliday}}, \bibinfo {author} {\bibfnamefont {S.~T.}\ \bibnamefont
  {Miller}}, \bibinfo {author} {\bibfnamefont {S.}~\bibnamefont {Sinha}}, \
  and\ \bibinfo {author} {\bibfnamefont {W.~L.}\ \bibnamefont {Ditto}},\
  }\bibfield  {title} {\enquote {\bibinfo {title} {Physics-enhanced neural
  networks learn order and chaos},}\ }\href@noop {} {\bibfield  {journal}
  {\bibinfo  {journal} {Phys. Rev. E}\ }\textbf {\bibinfo {volume} {101}},\
  \bibinfo {pages} {062207} (\bibinfo {year} {2020})}\BibitemShut {NoStop}%
\bibitem [{\citenamefont {Miller}\ \emph {et~al.}(2020)\citenamefont {Miller},
  \citenamefont {Lindner}, \citenamefont {Choudhary}, \citenamefont {Sinha},\
  and\ \citenamefont {Ditto}}]{sudhe3}%
  \BibitemOpen
  \bibfield  {author} {\bibinfo {author} {\bibfnamefont {S.~T.}\ \bibnamefont
  {Miller}}, \bibinfo {author} {\bibfnamefont {J.~F.}\ \bibnamefont {Lindner}},
  \bibinfo {author} {\bibfnamefont {A.}~\bibnamefont {Choudhary}}, \bibinfo
  {author} {\bibfnamefont {S.}~\bibnamefont {Sinha}}, \ and\ \bibinfo {author}
  {\bibfnamefont {W.~L.}\ \bibnamefont {Ditto}},\ }\bibfield  {title} {\enquote
  {\bibinfo {title} {The scaling of physics-informed machine learning with data
  and dimensions},}\ }\href@noop {} {\bibfield  {journal} {\bibinfo  {journal}
  {Chaos, Solitons \& Fractals: X}\ }\textbf {\bibinfo {volume} {5}},\ \bibinfo
  {pages} {100046} (\bibinfo {year} {2020})}\BibitemShut {NoStop}%
\bibitem [{\citenamefont {Mukhopadhyay}\ and\ \citenamefont
  {Banerjee}(2020)}]{santo1}%
  \BibitemOpen
  \bibfield  {author} {\bibinfo {author} {\bibfnamefont {S.}~\bibnamefont
  {Mukhopadhyay}}\ and\ \bibinfo {author} {\bibfnamefont {S.}~\bibnamefont
  {Banerjee}},\ }\bibfield  {title} {\enquote {\bibinfo {title} {Learning
  dynamical systems in noise using convolutional neural networks},}\
  }\href@noop {} {\bibfield  {journal} {\bibinfo  {journal} {Chaos}\ }\textbf
  {\bibinfo {volume} {30}},\ \bibinfo {pages} {103125} (\bibinfo {year}
  {2020})}\BibitemShut {NoStop}%
\bibitem [{\citenamefont {Pathak}\ \emph {et~al.}(2017)\citenamefont {Pathak},
  \citenamefont {Lu}, \citenamefont {Hunt}, \citenamefont {Girvan},\ and\
  \citenamefont {Ott}}]{Pathak2017}%
  \BibitemOpen
  \bibfield  {author} {\bibinfo {author} {\bibfnamefont {J.}~\bibnamefont
  {Pathak}}, \bibinfo {author} {\bibfnamefont {Z.}~\bibnamefont {Lu}}, \bibinfo
  {author} {\bibfnamefont {B.~R.}\ \bibnamefont {Hunt}}, \bibinfo {author}
  {\bibfnamefont {M.}~\bibnamefont {Girvan}}, \ and\ \bibinfo {author}
  {\bibfnamefont {E.}~\bibnamefont {Ott}},\ }\bibfield  {title} {\enquote
  {\bibinfo {title} {Using machine learning to replicate chaotic attractors and
  calculate {Lyapunov} exponents from data},}\ }\href@noop {} {\bibfield
  {journal} {\bibinfo  {journal} {Chaos}\ }\textbf {\bibinfo {volume} {27}},\
  \bibinfo {pages} {121102} (\bibinfo {year} {2017})}\BibitemShut {NoStop}%
\bibitem [{\citenamefont {Amil}, \citenamefont {Soriano},\ and\ \citenamefont
  {Masoller}(2019)}]{Amil2019}%
  \BibitemOpen
  \bibfield  {author} {\bibinfo {author} {\bibfnamefont {P.}~\bibnamefont
  {Amil}}, \bibinfo {author} {\bibfnamefont {M.~C.}\ \bibnamefont {Soriano}}, \
  and\ \bibinfo {author} {\bibfnamefont {C.}~\bibnamefont {Masoller}},\
  }\bibfield  {title} {\enquote {\bibinfo {title} {Machine learning algorithms
  for predicting the amplitude of chaotic laser pulses},}\ }\href@noop {}
  {\bibfield  {journal} {\bibinfo  {journal} {Chaos}\ }\textbf {\bibinfo
  {volume} {29}},\ \bibinfo {pages} {113111} (\bibinfo {year}
  {2019})}\BibitemShut {NoStop}%
\bibitem [{\citenamefont {Zhu}, \citenamefont {Ma},\ and\ \citenamefont
  {Lin}(2019)}]{zhu2019}%
  \BibitemOpen
  \bibfield  {author} {\bibinfo {author} {\bibfnamefont {Q.}~\bibnamefont
  {Zhu}}, \bibinfo {author} {\bibfnamefont {H.}~\bibnamefont {Ma}}, \ and\
  \bibinfo {author} {\bibfnamefont {W.}~\bibnamefont {Lin}},\ }\bibfield
  {title} {\enquote {\bibinfo {title} {Detecting unstable periodic orbits based
  only on time series: {When} adaptive delayed feedback control meets reservoir
  computing},}\ }\href@noop {} {\bibfield  {journal} {\bibinfo  {journal}
  {Chaos}\ }\textbf {\bibinfo {volume} {29}},\ \bibinfo {pages} {093125}
  (\bibinfo {year} {2019})}\BibitemShut {NoStop}%
\bibitem [{\citenamefont {Krishnagopal}\ \emph {et~al.}(2020)\citenamefont
  {Krishnagopal}, \citenamefont {Girvan}, \citenamefont {Ott},\ and\
  \citenamefont {Hunt}}]{Krishnagopal2020}%
  \BibitemOpen
  \bibfield  {author} {\bibinfo {author} {\bibfnamefont {S.}~\bibnamefont
  {Krishnagopal}}, \bibinfo {author} {\bibfnamefont {M.}~\bibnamefont
  {Girvan}}, \bibinfo {author} {\bibfnamefont {E.}~\bibnamefont {Ott}}, \ and\
  \bibinfo {author} {\bibfnamefont {B.~R.}\ \bibnamefont {Hunt}},\ }\bibfield
  {title} {\enquote {\bibinfo {title} {Separation of chaotic signals by
  reservoir computing},}\ }\href@noop {} {\bibfield  {journal} {\bibinfo
  {journal} {Chaos}\ }\textbf {\bibinfo {volume} {30}},\ \bibinfo {pages}
  {023123} (\bibinfo {year} {2020})}\BibitemShut {NoStop}%
\bibitem [{\citenamefont {Panday}\ \emph {et~al.}(2021)\citenamefont {Panday},
  \citenamefont {Lee}, \citenamefont {Dutta},\ and\ \citenamefont
  {Jalan}}]{Panday2021}%
  \BibitemOpen
  \bibfield  {author} {\bibinfo {author} {\bibfnamefont {A.}~\bibnamefont
  {Panday}}, \bibinfo {author} {\bibfnamefont {W.~S.}\ \bibnamefont {Lee}},
  \bibinfo {author} {\bibfnamefont {S.}~\bibnamefont {Dutta}}, \ and\ \bibinfo
  {author} {\bibfnamefont {S.}~\bibnamefont {Jalan}},\ }\bibfield  {title}
  {\enquote {\bibinfo {title} {Machine learning assisted network classification
  from symbolic time-series},}\ }\href@noop {} {\bibfield  {journal} {\bibinfo
  {journal} {Chaos}\ }\textbf {\bibinfo {volume} {31}},\ \bibinfo {pages}
  {031106} (\bibinfo {year} {2021})}\BibitemShut {NoStop}%
\bibitem [{\citenamefont {Barmparis}\ \emph {et~al.}(2020)\citenamefont
  {Barmparis}, \citenamefont {Neofotistos}, \citenamefont {Mattheakis},
  \citenamefont {Hizanidis}, \citenamefont {Tsironis},\ and\ \citenamefont
  {Kaxiras}}]{BARMPARIS2020}%
  \BibitemOpen
  \bibfield  {author} {\bibinfo {author} {\bibfnamefont {G.}~\bibnamefont
  {Barmparis}}, \bibinfo {author} {\bibfnamefont {G.}~\bibnamefont
  {Neofotistos}}, \bibinfo {author} {\bibfnamefont {M.}~\bibnamefont
  {Mattheakis}}, \bibinfo {author} {\bibfnamefont {J.}~\bibnamefont
  {Hizanidis}}, \bibinfo {author} {\bibfnamefont {G.}~\bibnamefont {Tsironis}},
  \ and\ \bibinfo {author} {\bibfnamefont {E.}~\bibnamefont {Kaxiras}},\
  }\bibfield  {title} {\enquote {\bibinfo {title} {Robust prediction of complex
  spatiotemporal states through machine learning with sparse sensing},}\
  }\href@noop {} {\bibfield  {journal} {\bibinfo  {journal} {Phys. Lett. A}\
  }\textbf {\bibinfo {volume} {384}},\ \bibinfo {pages} {126300} (\bibinfo
  {year} {2020})}\BibitemShut {NoStop}%
\bibitem [{\citenamefont {Ganaie}\ \emph {et~al.}(2020)\citenamefont {Ganaie},
  \citenamefont {Ghosh}, \citenamefont {Mendola}, \citenamefont {Tanveer},\
  and\ \citenamefont {Jalan}}]{ganaie2020}%
  \BibitemOpen
  \bibfield  {author} {\bibinfo {author} {\bibfnamefont {M.~A.}\ \bibnamefont
  {Ganaie}}, \bibinfo {author} {\bibfnamefont {S.}~\bibnamefont {Ghosh}},
  \bibinfo {author} {\bibfnamefont {N.}~\bibnamefont {Mendola}}, \bibinfo
  {author} {\bibfnamefont {M.}~\bibnamefont {Tanveer}}, \ and\ \bibinfo
  {author} {\bibfnamefont {S.}~\bibnamefont {Jalan}},\ }\bibfield  {title}
  {\enquote {\bibinfo {title} {Identification of chimera using machine
  learning},}\ }\href@noop {} {\bibfield  {journal} {\bibinfo  {journal}
  {Chaos}\ }\textbf {\bibinfo {volume} {30}},\ \bibinfo {pages} {063128}
  (\bibinfo {year} {2020})}\BibitemShut {NoStop}%
\bibitem [{\citenamefont {Kushwaha}\ \emph {et~al.}(2021)\citenamefont
  {Kushwaha}, \citenamefont {Mendola}, \citenamefont {Ghosh}, \citenamefont
  {Kachhvah},\ and\ \citenamefont {Jalan}}]{kushwaha2020}%
  \BibitemOpen
  \bibfield  {author} {\bibinfo {author} {\bibfnamefont {N.}~\bibnamefont
  {Kushwaha}}, \bibinfo {author} {\bibfnamefont {N.~K.}\ \bibnamefont
  {Mendola}}, \bibinfo {author} {\bibfnamefont {S.}~\bibnamefont {Ghosh}},
  \bibinfo {author} {\bibfnamefont {A.~D.}\ \bibnamefont {Kachhvah}}, \ and\
  \bibinfo {author} {\bibfnamefont {S.}~\bibnamefont {Jalan}},\ }\bibfield
  {title} {\enquote {\bibinfo {title} {Machine learning assisted chimera and
  solitary states in networks},}\ }\href@noop {} {\bibfield  {journal}
  {\bibinfo  {journal} {Front. Phys.}\ }\textbf {\bibinfo {volume} {9}},\
  \bibinfo {pages} {147} (\bibinfo {year} {2021})}\BibitemShut {NoStop}%
\bibitem [{\citenamefont {Lellep}\ \emph {et~al.}(2020)\citenamefont {Lellep},
  \citenamefont {Prexl}, \citenamefont {Linkmann},\ and\ \citenamefont
  {Eckhardt}}]{lellep2020}%
  \BibitemOpen
  \bibfield  {author} {\bibinfo {author} {\bibfnamefont {M.}~\bibnamefont
  {Lellep}}, \bibinfo {author} {\bibfnamefont {J.}~\bibnamefont {Prexl}},
  \bibinfo {author} {\bibfnamefont {M.}~\bibnamefont {Linkmann}}, \ and\
  \bibinfo {author} {\bibfnamefont {B.}~\bibnamefont {Eckhardt}},\ }\bibfield
  {title} {\enquote {\bibinfo {title} {Using machine learning to predict
  extreme events in the hénon map},}\ }\href@noop {} {\bibfield  {journal}
  {\bibinfo  {journal} {Chaos}\ }\textbf {\bibinfo {volume} {30}},\ \bibinfo
  {pages} {013113} (\bibinfo {year} {2020})}\BibitemShut {NoStop}%
\bibitem [{\citenamefont {Meiyazhagan}, \citenamefont {Sudharsan},\ and\
  \citenamefont {Senthilvelan}(2021)}]{meiysudha1}%
  \BibitemOpen
  \bibfield  {author} {\bibinfo {author} {\bibfnamefont {J.}~\bibnamefont
  {Meiyazhagan}}, \bibinfo {author} {\bibfnamefont {S.}~\bibnamefont
  {Sudharsan}}, \ and\ \bibinfo {author} {\bibfnamefont {M.}~\bibnamefont
  {Senthilvelan}},\ }\bibfield  {title} {\enquote {\bibinfo {title} {Model-free
  prediction of emergence of extreme events in a parametrically driven
  nonlinear dynamical system by deep learning},}\ }\href@noop {} {\bibfield
  {journal} {\bibinfo  {journal} {Eur. Phys. J. B}\ }\textbf {\bibinfo {volume}
  {94}},\ \bibinfo {pages} {1--13} (\bibinfo {year} {2021})}\BibitemShut
  {NoStop}%
\bibitem [{\citenamefont {Chowdhury}\ \emph {et~al.}(2021)\citenamefont
  {Chowdhury}, \citenamefont {Ray}, \citenamefont {Mishra},\ and\ \citenamefont
  {Ghosh}}]{dibak1}%
  \BibitemOpen
  \bibfield  {author} {\bibinfo {author} {\bibfnamefont {S.~N.}\ \bibnamefont
  {Chowdhury}}, \bibinfo {author} {\bibfnamefont {A.}~\bibnamefont {Ray}},
  \bibinfo {author} {\bibfnamefont {A.}~\bibnamefont {Mishra}}, \ and\ \bibinfo
  {author} {\bibfnamefont {D.}~\bibnamefont {Ghosh}},\ }\bibfield  {title}
  {\enquote {\bibinfo {title} {Extreme events in globally coupled chaotic
  maps},}\ }\href@noop {} {\bibfield  {journal} {\bibinfo  {journal} {J. Phys.
  Complex.}\ }\textbf {\bibinfo {volume} {2}},\ \bibinfo {pages} {035021}
  (\bibinfo {year} {2021})}\BibitemShut {NoStop}%
\bibitem [{\citenamefont {Ray}, \citenamefont {Chakraborty},\ and\
  \citenamefont {Ghosh}(2021)}]{ray2021optimized}%
  \BibitemOpen
  \bibfield  {author} {\bibinfo {author} {\bibfnamefont {A.}~\bibnamefont
  {Ray}}, \bibinfo {author} {\bibfnamefont {T.}~\bibnamefont {Chakraborty}}, \
  and\ \bibinfo {author} {\bibfnamefont {D.}~\bibnamefont {Ghosh}},\ }\bibfield
   {title} {\enquote {\bibinfo {title} {Optimized ensemble deep learning
  framework for scalable forecasting of dynamics containing extreme events},}\
  }\href@noop {} {\bibfield  {journal} {\bibinfo  {journal} {arXiv}\ }\textbf
  {\bibinfo {volume} {2106.08968}} (\bibinfo {year} {2021})}\BibitemShut
  {NoStop}%
\bibitem [{\citenamefont {Meiyazhagan}\ \emph {et~al.}(2022)\citenamefont
  {Meiyazhagan}, \citenamefont {Sudharsan}, \citenamefont {Venkatesan},\ and\
  \citenamefont {Senthilvelan}}]{meiysudha2}%
  \BibitemOpen
  \bibfield  {author} {\bibinfo {author} {\bibfnamefont {J.}~\bibnamefont
  {Meiyazhagan}}, \bibinfo {author} {\bibfnamefont {S.}~\bibnamefont
  {Sudharsan}}, \bibinfo {author} {\bibfnamefont {A.}~\bibnamefont
  {Venkatesan}}, \ and\ \bibinfo {author} {\bibfnamefont {M.}~\bibnamefont
  {Senthilvelan}},\ }\bibfield  {title} {\enquote {\bibinfo {title} {Prediction
  of occurrence of extreme events using machine learning},}\ }\href@noop {}
  {\bibfield  {journal} {\bibinfo  {journal} {Eur. Phys. J. Plus}\ }\textbf
  {\bibinfo {volume} {137}},\ \bibinfo {pages} {1--20} (\bibinfo {year}
  {2022})}\BibitemShut {NoStop}%
\bibitem [{\citenamefont {Raissi}, \citenamefont {Perdikaris},\ and\
  \citenamefont {Karniadakis}(2019{\natexlab{a}})}]{raissi2019physics}%
  \BibitemOpen
  \bibfield  {author} {\bibinfo {author} {\bibfnamefont {M.}~\bibnamefont
  {Raissi}}, \bibinfo {author} {\bibfnamefont {P.}~\bibnamefont {Perdikaris}},
  \ and\ \bibinfo {author} {\bibfnamefont {G.~E.}\ \bibnamefont
  {Karniadakis}},\ }\bibfield  {title} {\enquote {\bibinfo {title}
  {Physics-informed neural networks: A deep learning framework for solving
  forward and inverse problems involving nonlinear partial differential
  equations},}\ }\href@noop {} {\bibfield  {journal} {\bibinfo  {journal} {J.
  Comput. Phys.}\ }\textbf {\bibinfo {volume} {378}},\ \bibinfo {pages}
  {686--707} (\bibinfo {year} {2019}{\natexlab{a}})}\BibitemShut {NoStop}%
\bibitem [{\citenamefont {Pu}, \citenamefont {Peng},\ and\ \citenamefont
  {Chen}(2021)}]{pu2021data}%
  \BibitemOpen
  \bibfield  {author} {\bibinfo {author} {\bibfnamefont {J.}~\bibnamefont
  {Pu}}, \bibinfo {author} {\bibfnamefont {W.}~\bibnamefont {Peng}}, \ and\
  \bibinfo {author} {\bibfnamefont {Y.}~\bibnamefont {Chen}},\ }\bibfield
  {title} {\enquote {\bibinfo {title} {The data-driven localized wave solutions
  of the derivative nonlinear {Schr{\"o}dinger} equation by using improved
  {PINN} approach},}\ }\href@noop {} {\bibfield  {journal} {\bibinfo  {journal}
  {Wave Motion}\ }\textbf {\bibinfo {volume} {107}},\ \bibinfo {pages} {102823}
  (\bibinfo {year} {2021})}\BibitemShut {NoStop}%
\bibitem [{\citenamefont {Wang}\ and\ \citenamefont
  {Yan}(2021{\natexlab{a}})}]{wang2021data}%
  \BibitemOpen
  \bibfield  {author} {\bibinfo {author} {\bibfnamefont {L.}~\bibnamefont
  {Wang}}\ and\ \bibinfo {author} {\bibfnamefont {Z.}~\bibnamefont {Yan}},\
  }\bibfield  {title} {\enquote {\bibinfo {title} {Data-driven rogue waves and
  parameter discovery in the defocusing nonlinear {Schr{\"o}dinger} equation
  with a potential using the {PINN} deep learning},}\ }\href@noop {} {\bibfield
   {journal} {\bibinfo  {journal} {Phys. Lett. A}\ }\textbf {\bibinfo {volume}
  {404}},\ \bibinfo {pages} {127408} (\bibinfo {year}
  {2021}{\natexlab{a}})}\BibitemShut {NoStop}%
\bibitem [{\citenamefont {Zhou}\ and\ \citenamefont
  {Yan}(2021{\natexlab{a}})}]{zhou2021deep}%
  \BibitemOpen
  \bibfield  {author} {\bibinfo {author} {\bibfnamefont {Z.}~\bibnamefont
  {Zhou}}\ and\ \bibinfo {author} {\bibfnamefont {Z.}~\bibnamefont {Yan}},\
  }\bibfield  {title} {\enquote {\bibinfo {title} {Deep learning neural
  networks for the third-order nonlinear {Schr{\"o}dinger} equation:
  {Solitons}, breathers, and rogue waves},}\ }\href@noop {} {\bibfield
  {journal} {\bibinfo  {journal} {arXiv}\ }\textbf {\bibinfo {volume}
  {2104.14809}} (\bibinfo {year} {2021}{\natexlab{a}})}\BibitemShut {NoStop}%
\bibitem [{\citenamefont {Wang}\ and\ \citenamefont
  {Yan}(2021{\natexlab{b}})}]{wang2021data2}%
  \BibitemOpen
  \bibfield  {author} {\bibinfo {author} {\bibfnamefont {L.}~\bibnamefont
  {Wang}}\ and\ \bibinfo {author} {\bibfnamefont {Z.}~\bibnamefont {Yan}},\
  }\bibfield  {title} {\enquote {\bibinfo {title} {Data-driven peakon and
  periodic peakon solutions and parameter discovery of some nonlinear
  dispersive equations via deep learning},}\ }\href@noop {} {\bibfield
  {journal} {\bibinfo  {journal} {Physica D}\ }\textbf {\bibinfo {volume}
  {428}},\ \bibinfo {pages} {133037} (\bibinfo {year}
  {2021}{\natexlab{b}})}\BibitemShut {NoStop}%
\bibitem [{\citenamefont {Mo}, \citenamefont {Ling},\ and\ \citenamefont
  {Zeng}(2022)}]{mo2022data}%
  \BibitemOpen
  \bibfield  {author} {\bibinfo {author} {\bibfnamefont {Y.}~\bibnamefont
  {Mo}}, \bibinfo {author} {\bibfnamefont {L.}~\bibnamefont {Ling}}, \ and\
  \bibinfo {author} {\bibfnamefont {D.}~\bibnamefont {Zeng}},\ }\bibfield
  {title} {\enquote {\bibinfo {title} {Data-driven vector soliton solutions of
  coupled nonlinear {Schr{\"o}dinger} equation using a deep learning
  algorithm},}\ }\href@noop {} {\bibfield  {journal} {\bibinfo  {journal}
  {Phys. Lett. A}\ }\textbf {\bibinfo {volume} {421}},\ \bibinfo {pages}
  {127739} (\bibinfo {year} {2022})}\BibitemShut {NoStop}%
\bibitem [{\citenamefont {Zhou}\ and\ \citenamefont
  {Yan}(2021{\natexlab{b}})}]{zhou2021solving}%
  \BibitemOpen
  \bibfield  {author} {\bibinfo {author} {\bibfnamefont {Z.}~\bibnamefont
  {Zhou}}\ and\ \bibinfo {author} {\bibfnamefont {Z.}~\bibnamefont {Yan}},\
  }\bibfield  {title} {\enquote {\bibinfo {title} {Solving forward and inverse
  problems of the logarithmic nonlinear {Schr{\"o}dinger} equation with
  $\mathcal{PT}$-symmetric harmonic potential via deep learning},}\ }\href@noop
  {} {\bibfield  {journal} {\bibinfo  {journal} {Phys. Lett. A}\ }\textbf
  {\bibinfo {volume} {387}},\ \bibinfo {pages} {127010} (\bibinfo {year}
  {2021}{\natexlab{b}})}\BibitemShut {NoStop}%
\bibitem [{\citenamefont {Li}\ and\ \citenamefont {Li}(2021)}]{li2021solving}%
  \BibitemOpen
  \bibfield  {author} {\bibinfo {author} {\bibfnamefont {J.}~\bibnamefont
  {Li}}\ and\ \bibinfo {author} {\bibfnamefont {B.}~\bibnamefont {Li}},\
  }\bibfield  {title} {\enquote {\bibinfo {title} {Solving forward and inverse
  problems of the nonlinear {Schr{\"o}dinger} equation with the generalized
  $\mathcal{PT}$-symmetric {Scarf-II} potential via {PINN} deep learning},}\
  }\href@noop {} {\bibfield  {journal} {\bibinfo  {journal} {Commun. Theor.
  Phys.}\ }\textbf {\bibinfo {volume} {73}},\ \bibinfo {pages} {125001}
  (\bibinfo {year} {2021})}\BibitemShut {NoStop}%
\bibitem [{\citenamefont {Raissi}, \citenamefont {Perdikaris},\ and\
  \citenamefont {Karniadakis}(2019{\natexlab{b}})}]{raissi2019}%
  \BibitemOpen
  \bibfield  {author} {\bibinfo {author} {\bibfnamefont {M.}~\bibnamefont
  {Raissi}}, \bibinfo {author} {\bibfnamefont {P.}~\bibnamefont {Perdikaris}},
  \ and\ \bibinfo {author} {\bibfnamefont {G.~E.}\ \bibnamefont
  {Karniadakis}},\ }\bibfield  {title} {\enquote {\bibinfo {title}
  {Physics-informed neural networks: A deep learning framework for solving
  forward and inverse problems involving nonlinear partial differential
  equations},}\ }\href@noop {} {\bibfield  {journal} {\bibinfo  {journal} {J.
  Comput. Phys.}\ }\textbf {\bibinfo {volume} {378}},\ \bibinfo {pages}
  {686--707} (\bibinfo {year} {2019}{\natexlab{b}})}\BibitemShut {NoStop}%
\bibitem [{\citenamefont {Baydin}\ \emph {et~al.}(2018)\citenamefont {Baydin},
  \citenamefont {Pearlmutter}, \citenamefont {Radul},\ and\ \citenamefont
  {Siskind}}]{baydin2018}%
  \BibitemOpen
  \bibfield  {author} {\bibinfo {author} {\bibfnamefont {A.~G.}\ \bibnamefont
  {Baydin}}, \bibinfo {author} {\bibfnamefont {B.~A.}\ \bibnamefont
  {Pearlmutter}}, \bibinfo {author} {\bibfnamefont {A.~A.}\ \bibnamefont
  {Radul}}, \ and\ \bibinfo {author} {\bibfnamefont {J.~M.}\ \bibnamefont
  {Siskind}},\ }\bibfield  {title} {\enquote {\bibinfo {title} {Automatic
  differentiation in machine learning: a survey},}\ }\href@noop {} {\bibfield
  {journal} {\bibinfo  {journal} {J. Mach. Learn. Res.}\ }\textbf {\bibinfo
  {volume} {18}} (\bibinfo {year} {2018})}\BibitemShut {NoStop}%
\bibitem [{\citenamefont {Margossian}(2019)}]{margossian2019review}%
  \BibitemOpen
  \bibfield  {author} {\bibinfo {author} {\bibfnamefont {C.~C.}\ \bibnamefont
  {Margossian}},\ }\bibfield  {title} {\enquote {\bibinfo {title} {A review of
  automatic differentiation and its efficient implementation},}\ }\href@noop {}
  {\bibfield  {journal} {\bibinfo  {journal} {Wiley interdisciplinary reviews:
  data mining and knowledge discovery}\ }\textbf {\bibinfo {volume} {9}},\
  \bibinfo {pages} {e1305} (\bibinfo {year} {2019})}\BibitemShut {NoStop}%
\bibitem [{\citenamefont {Rumelhart}, \citenamefont {Hinton},\ and\
  \citenamefont {Williams}(1986)}]{rumelhart1986}%
  \BibitemOpen
  \bibfield  {author} {\bibinfo {author} {\bibfnamefont {D.~E.}\ \bibnamefont
  {Rumelhart}}, \bibinfo {author} {\bibfnamefont {G.~E.}\ \bibnamefont
  {Hinton}}, \ and\ \bibinfo {author} {\bibfnamefont {R.~J.}\ \bibnamefont
  {Williams}},\ }\bibfield  {title} {\enquote {\bibinfo {title} {Learning
  representations by back-propagating errors},}\ }\href@noop {} {\bibfield
  {journal} {\bibinfo  {journal} {Nature}\ }\textbf {\bibinfo {volume} {323}},\
  \bibinfo {pages} {533--536} (\bibinfo {year} {1986})}\BibitemShut {NoStop}%
\bibitem [{\citenamefont {Abadi}\ \emph {et~al.}(2016)\citenamefont {Abadi},
  \citenamefont {Barham}, \citenamefont {Chen}, \citenamefont {Chen},
  \citenamefont {Davis}, \citenamefont {Dean}, \citenamefont {Devin},
  \citenamefont {Ghemawat}, \citenamefont {Irving}, \citenamefont {Isard} \emph
  {et~al.}}]{abadi2016}%
  \BibitemOpen
  \bibfield  {author} {\bibinfo {author} {\bibfnamefont {M.}~\bibnamefont
  {Abadi}}, \bibinfo {author} {\bibfnamefont {P.}~\bibnamefont {Barham}},
  \bibinfo {author} {\bibfnamefont {J.}~\bibnamefont {Chen}}, \bibinfo {author}
  {\bibfnamefont {Z.}~\bibnamefont {Chen}}, \bibinfo {author} {\bibfnamefont
  {A.}~\bibnamefont {Davis}}, \bibinfo {author} {\bibfnamefont
  {J.}~\bibnamefont {Dean}}, \bibinfo {author} {\bibfnamefont {M.}~\bibnamefont
  {Devin}}, \bibinfo {author} {\bibfnamefont {S.}~\bibnamefont {Ghemawat}},
  \bibinfo {author} {\bibfnamefont {G.}~\bibnamefont {Irving}}, \bibinfo
  {author} {\bibfnamefont {M.}~\bibnamefont {Isard}},  \emph {et~al.},\
  }\bibfield  {title} {\enquote {\bibinfo {title} {Tensorflow: A system for
  large-scale machine learning},}\ }in\ \href@noop {} {\emph {\bibinfo
  {booktitle} {12th $\{$USENIX$\}$ symposium on operating systems design and
  implementation ($\{$OSDI$\}$ 16)}}}\ (\bibinfo {year} {2016})\ pp.\ \bibinfo
  {pages} {265--283}\BibitemShut {NoStop}%
\bibitem [{\citenamefont {Stein}(1987)}]{stein1987}%
  \BibitemOpen
  \bibfield  {author} {\bibinfo {author} {\bibfnamefont {M.}~\bibnamefont
  {Stein}},\ }\bibfield  {title} {\enquote {\bibinfo {title} {Large sample
  properties of simulations using {Latin} hypercube sampling},}\ }\href@noop {}
  {\bibfield  {journal} {\bibinfo  {journal} {Technometrics}\ }\textbf
  {\bibinfo {volume} {29}},\ \bibinfo {pages} {143--151} (\bibinfo {year}
  {1987})}\BibitemShut {NoStop}%
\bibitem [{\citenamefont {Liu}\ and\ \citenamefont {Nocedal}(1989)}]{liu1989}%
  \BibitemOpen
  \bibfield  {author} {\bibinfo {author} {\bibfnamefont {D.~C.}\ \bibnamefont
  {Liu}}\ and\ \bibinfo {author} {\bibfnamefont {J.}~\bibnamefont {Nocedal}},\
  }\bibfield  {title} {\enquote {\bibinfo {title} {On the limited memory {BFGS}
  method for large scale optimization},}\ }\href@noop {} {\bibfield  {journal}
  {\bibinfo  {journal} {Math. Program.}\ }\textbf {\bibinfo {volume} {45}},\
  \bibinfo {pages} {503--528} (\bibinfo {year} {1989})}\BibitemShut {NoStop}%
\bibitem [{\citenamefont {Yang}(2010)}]{yang2010}%
  \BibitemOpen
  \bibfield  {author} {\bibinfo {author} {\bibfnamefont {J.}~\bibnamefont
  {Yang}},\ }\href@noop {} {\emph {\bibinfo {title} {Nonlinear waves in
  integrable and nonintegrable systems}}}\ (\bibinfo  {publisher} {SIAM},\
  \bibinfo {year} {2010})\BibitemShut {NoStop}%
\bibitem [{\citenamefont {Hu}\ \emph {et~al.}(2011)\citenamefont {Hu},
  \citenamefont {Ma}, \citenamefont {Lu}, \citenamefont {Yang}, \citenamefont
  {Zheng},\ and\ \citenamefont {Hu}}]{hu2011}%
  \BibitemOpen
  \bibfield  {author} {\bibinfo {author} {\bibfnamefont {S.}~\bibnamefont
  {Hu}}, \bibinfo {author} {\bibfnamefont {X.}~\bibnamefont {Ma}}, \bibinfo
  {author} {\bibfnamefont {D.}~\bibnamefont {Lu}}, \bibinfo {author}
  {\bibfnamefont {Z.}~\bibnamefont {Yang}}, \bibinfo {author} {\bibfnamefont
  {Y.}~\bibnamefont {Zheng}}, \ and\ \bibinfo {author} {\bibfnamefont
  {W.}~\bibnamefont {Hu}},\ }\bibfield  {title} {\enquote {\bibinfo {title}
  {Solitons supported by complex $\mathcal{PT}$-symmetric {Gaussian}
  potentials},}\ }\href@noop {} {\bibfield  {journal} {\bibinfo  {journal}
  {Phys. Rev. A}\ }\textbf {\bibinfo {volume} {84}},\ \bibinfo {pages} {043818}
  (\bibinfo {year} {2011})}\BibitemShut {NoStop}%
\bibitem [{\citenamefont {Midya}\ and\ \citenamefont
  {Roychoudhury}(2013)}]{midya2013}%
  \BibitemOpen
  \bibfield  {author} {\bibinfo {author} {\bibfnamefont {B.}~\bibnamefont
  {Midya}}\ and\ \bibinfo {author} {\bibfnamefont {R.}~\bibnamefont
  {Roychoudhury}},\ }\bibfield  {title} {\enquote {\bibinfo {title} {Nonlinear
  localized modes in $\mathcal{PT}$-symmetric {Rosen-Morse} potential wells},}\
  }\href@noop {} {\bibfield  {journal} {\bibinfo  {journal} {Phys. Rev. A}\
  }\textbf {\bibinfo {volume} {87}},\ \bibinfo {pages} {045803} (\bibinfo
  {year} {2013})}\BibitemShut {NoStop}%
\end{thebibliography}%
	
\end{document}